\documentclass{aa}  

\usepackage{graphicx}
\usepackage{txfonts}
\usepackage[]{hyperref}  
\usepackage{threeparttable}
\usepackage{subcaption}
\usepackage{graphicx}
\usepackage{txfonts}

\hypersetup{
  colorlinks=true,  
  urlcolor=blue,    
  linkcolor=red,
  citecolor=blue     
}

\usepackage[normalem]{ulem}

\begin{document} 

   \title{Sun-as-a-star analysis of simulated solar flares}

   \author{H. C. Yu\inst{1,2}
          , J. Hong\inst{3,1}
          \and
          M. D. Ding\inst{1,2}
          }

   \institute{
   School of Astronomy and Space Science, Nanjing University, 163 Xianlin Road, Nanjing 210023, PR China\\
   \email{dmd@nju.edu.cn}
         \and
             Key Laboratory for Modern Astronomy and Astrophysics (Nanjing University), Ministry of Education, Nanjing 210023, 163 Xianlin Road, PR China
        \and 
            Institute for Solar Physics, Dept. of Astronomy, Stockholm University, AlbaNova University Centre, SE-106 91 Stockholm, Sweden\\
            \email{jie.hong@astro.su.se}
             }

   \date{}

  \abstract
   {Stellar flares have an impact on habitable planets. To relate the observations of the Sun with those of stars, one needs to use a Sun-as-a-star analysis, that is, to degrade the resolution of the Sun to a single point. With the data of the Sun-as-a-star observations, a simulation of solar flares is required to provide a systemic clue for the Sun-as-a-star study.}
   {We aim to explore how the Sun-as-a-star spectrum varies with the flare magnitude and location based on a grid of solar flare models.}
   {Using 1D radiative hydrodynamics modeling and multi-thread flare assumption, we obtained the spectrum of a typical flare with an enhancement of chromospheric lines.}
   {The Sun-as-a-star spectrum of the H$\alpha$ line shows enhanced and shifted components, which are highly dependent on the flare magnitude and location. The equivalent width $\Delta\mathrm{EW}$ is a good indicator of energy release. The bisector method can be used to diagnose the sign of the line-of-sight velocity in the flaring atmosphere. For both H$\alpha$ and H$\beta$ lines, the Sun-as-a-star spectrum of a limb flare tends to be wider and shows a dip in the line center. In particular, we propose two quantities to diagnose the magnitude and location of the stellar flares. Besides this, caution must be taken when calculating the radiation energy, since the astrophysical flux-to-energy conversion ratio is dependent on the flare location.}
   {}


   \maketitle

\section{Introduction}
        In order to search for habitable planets, it is necessary to learn the change in the interplanetary space environment that originated from the hosting star. A star with intense and frequent activities can strongly influence the space weather; such activities have a significant impact on the creation and maintenance of life on planets. Stellar flares are a common type of stellar activity, which release a large amount of energy in a short period \citep{2024Kowalski}. 
        Flares are thought to originate from magnetic reconnections and are sometimes accompanied by coronal mass ejections (CMEs) and energetic particles, which can hazardously change the space weather \citep{2007Pulkkinen,2003Tsurtani}. 
        During flares, energy is released from the magnetic field in the upper atmosphere \citep{1963Parker,2001Shibata,2011Shibata}, which is partly converted to a radiation enhancement observed in multi-spectral windows such as the radio, optical, ultraviolet, X-rays, and so on. For solar flares, the released energy is typically $10^{29}$ to $10^{33}$ erg. In other stars, the total energy can reach up to $10^{36}$ erg \citep{2012Maehara}. However, there is recent a debate about whether the total flare energy can be derived from the assumption of black-body radiation or hydrogen recombination continuum. The latter would lower the estimated total energy by one order of magnitude \citep{2024Simoes}.

    Since stars are too distant to be observed with sufficient spatial resolution, as the Sun has, determining the flare level and even its location on the stars is a significant challenge. Therefore, one usually resorts to observations and modeling of the Sun. Studying the Sun provides a bridge to understanding what could happen on other stars -- especially solar-type stars \citep{2013Notsu} -- since all transient events on the Sun may also occur in other types of stars \citep{1991Haisch}.

    To relate the observations of the Sun with those of stars, one has to degrade the resolution of the Sun to a single point, that is, integrate the observables over the full disk of the Sun. The method is called the Sun-as-a-star analysis (e.g. \cite{2004Woods,2011Ketzschmar,2024Pietrowc}). This method is widely used in determining the long-period variation of the total solar irradiance (TSI) \citep{2007Livingston,2023Criscuoli,2024Zills}. Today, the Sun-as-a-star method is also used for detecting short-period events such as solar flares and filament eruptions \citep{2022Otsu,2022Namekatac,2022Xu,2024Pietrow}. 

    For solar flares, the flaring region is too small compared with the whole solar disk, making it very difficult to directly see the flare signals from the optical spectra. A practical way to extract the flare signals is to study the Sun-as-a-star spectrum subtracted by the pre-flare one.
    \cite{2022Namekatac} calculated the pre-flare-subtracted H$\alpha$ spectra taken by the Solar Dynamics Doppler Imager, from which the velocity and line width are obtained. \cite{2022Otsu,2024Otsu} used the same instrument and method to analyze several flares, filament eruptions, and prominence eruptions. They suggest that the different characteristics in the Sun-as-a-star profiles can be evidence of different kinds of activities. 
 
    Furthermore, the result of \cite{2022Otsu} suggests that the H$\alpha$ line emission is related to the location of the flare, showing a center-to-limb variation (CLV). The CLV effect has been proven to be significant in the quiet-Sun spectrum \citep{2023Pietrowa,2024Pietrowc}. Such an effect is even more conspicuous when a flare occurs, in other words, two identical flares can result in very different Sun-as-a-star spectra if one is located at the disk center and the other is at the disk limb. By considering the CLV effect on flare observations, \cite{2006Woods} used the traditional limb-darkening formula to obtain the ratio of the TSI to the GOES X-ray flare energy as a function of the flare location. Although there have been some previous studies, the CLV effect in solar or stellar flares is still poorly understood and requires further in-depth research, as indicated by \cite{2024Notsu}. One may notice that the previous studies are mostly based on observations, and those based on simulations are still lacking. By comparison, simulations are more illustrative in revealing the CLV effect in dependence on key parameters like the flare magnitudes and locations, thus providing a clue to diagnosing stellar flares. 
    
    In this paper, we introduce a model to calculate the Sun-as-a-star spectrum of chromospheric lines during a flare. A specific parameter study on the Sun-as-a-star spectrum is presented. The two key parameters in our model include the flare magnitude and location. In Sect. \ref{method}, we introduce the model used in our work. We show the result in Sect. \ref{Result}, and we compare it to observations in Sect. \ref{comparing to observation}. Finally, we summarize our results in Sect. \ref{conclusion}.

\section{Method}
    \label{method}
    Observations with high spatial resolution have shown that flares are not composed of a single loop \citep{2001Aschwanden} but a series of loops. To be realistic, a multi-thread flare simulation is suggested \citep{2006Warren,2016Costa,2017Kowalski}. The principle of multi-thread simulation is that the foot points of flare loops brighten consecutively as the magnetic reconnection proceeds and all the brightened foot points contribute to the total flare radiation. We adopt such a model in this work. We first simulated the spectrum of a flaring loop, which is regarded as a single thread. Then, we calculated the multi-thread spectrum based on the result of the single thread.

    \subsection{Simulation of a single thread}
    \label{Flare Modeling}
    
    The \verb"RADYN" code is a one-dimensional radiative hydrodynamic code based on a plane-parallel atmosphere \citep{1992ApJ...397L..59C,1995ApJ...440L..29C,1997ApJ...481..500C,2002ApJ...572..626C}. It is now used more often to calculate the atmospheric response under flare conditions \citep{2015Allred}. 
    Solar flares are assumed to be caused by nonthermal electrons, which are produced by magnetic reconnection in the corona. The electrons lose their energy by interacting with ambient particles through coulomb collisions, leading to the heating of the atmosphere in the meantime. In our simulation, the initial atmosphere is based on the VAL3C model, with a 20 Mm flare loop. We only simulated one leg of a semicircular loop from the photosphere to the corona (10 Mm). A nonthermal electron beam is injected from the loop top and travels downward along the loop. For each simulation thread, the energy flux of the electron beam linearly increases with time in the first 10 s, maintains the maximum value for 10 s, and then linearly decreases for the next 10 s. After that, the atmosphere continues to evolve with no heating for the last 30 s. 
    The total simulation time for a single thread is thus 60 s, which is on the same order as in \cite{2016Costa}.
    We assume that the electrons follow a power-law distribution, with a fixed cut-off energy of $E_c=25$ keV and a fixed spectral index of $\delta=3$. These two parameters change the penetration depth of the beam, and we fixed their values to some typical ones similar to those seen in previous flare simulations \citep{2019ApJ...871...23K,2022Hong,2023Yu}. The peak energy flux $F_\mathrm{peak}$ is a key parameter in flare simulations. Here, we take ten values from $10^{10}\ \mathrm{erg}\ \mathrm{s^{-1}}\ \mathrm{cm^{-2}}$ to $10^{11}\ \mathrm{erg}\ \mathrm{s^{-1}}\ \mathrm{cm^{-2}}$ for $F_\mathrm{peak}$, so as to cover different energy magnitudes of flares (hereafter noted as 1 to 10). The quadricircular loop is 10 Mm in length and the initial atmosphere is adopted to be the quiet-Sun model with a loop-top temperature of 1 MK. 

    In the simulation, we calculated the emergent intensity for five positions on the solar disk denoted by the cosine of the heliocentric angle $\mu\equiv\cos\theta$. The values of $\mu$ are fixed at 0.047, 0.231, 0.500, 0.769, and 0.953, from the solar limb to the disk center. We ran the above single-thread simulation for 60 s and saved the snapshot every 1 s.  
    
    The spectral lines in \verb"RADYN" are all calculated with the assumption of complete frequency redistribution, which is inaccurate for strong resonance lines such as \ion{Ca}{II} K\&H. So, we only took the H$\alpha$ and H$\beta$ lines from \verb"RADYN" outputs. We then fed the simulation output to \verb"RH 1.5D" \citep{2015Pereira} to recalculate the spectral lines (assuming partial frequency redistribution) and continua with different flare locations $\mu$ and a different electron beam flux $F_\mathrm{peak}$. We took all the \ion{Ca}{II} K\&H lines and continua from here. Thus, for each value of $F_\mathrm{peak}$, we obtain a three-dimensional data matrix for the emergent intensity as $I(\mu,\lambda,t)$.

    \subsection{Multi-thread model}

    \begin{figure}[b]
    \centering
    \includegraphics[width=0.5\textwidth]{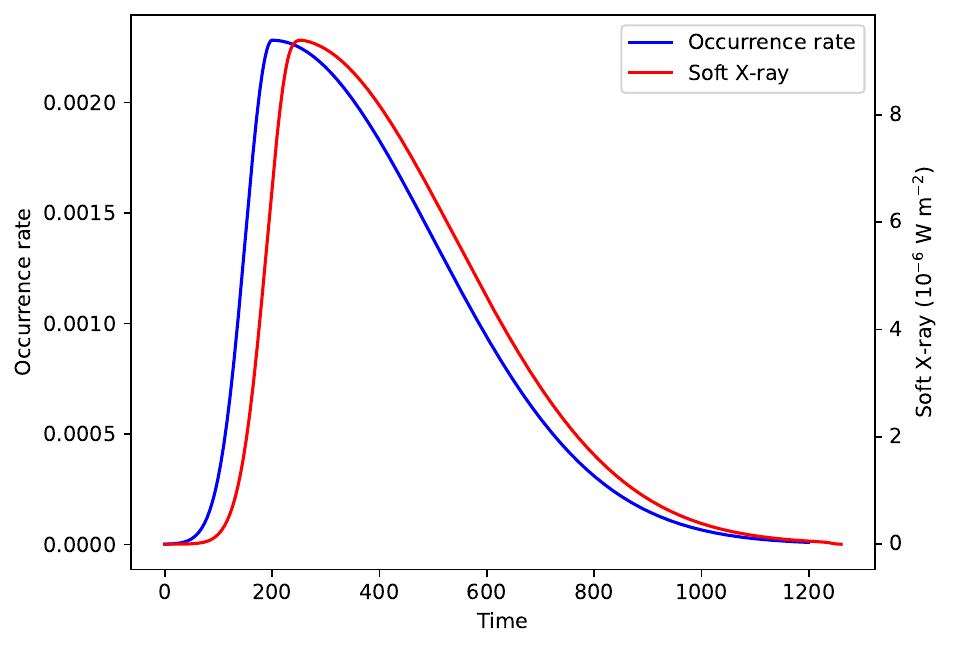}
    \caption{Occurrence rate $P(t)$ (blue) and GOES 1-8 \AA\ flux when $F_\mathrm{peak}=5$ (red) as a function of time. The X-ray synthetic method is introduced in Sect. \ref{activity indices}.}
    \label{rate}
    \end{figure}
  
    \cite{2016Qiu} proposed two modes of multi-thread simulation: amplitude modulation and frequency modulation. Amplitude modulation modifies the intensity of each thread, while the time interval between two consecutive threads is constant. Frequency modulation means that the threads are all the same, but their occurrence rate changes with time. For both modulation types, every single loop brightens independently and does not interact with the adjacent loop.
    
    Given a flare location $\mu$ and electron beam flux $F_\mathrm{peak}$, we used the multi-thread simulation with frequency modulation. At a given time $t$, the flare loops in the flare region are at different heating stages. We introduce an occurrence rate $P(t)$ to denote the gradient of the cumulative fraction of loops that have started flaring. Thus, $P(t)\mathrm{d}t$ is equal to the percentage of the flare loops that start being heated at the time from $t$ to $t+\mathrm{d}t$, with $\int_{0}^{T}P(t)\mathrm{d}t=1$, where $T$ is the total time when the heating of different threads is initiated in the multi-thread model. 
    The average intensity of the flare region is then expressed as
    \begin{equation}
    \begin{split}
     I_F(\mu,\lambda,t)=\int_{0}^{T}P(t')I(\mu,\lambda,t-t')\mathrm{d}t'.
    \end{split}
    \label{FM}
    \end{equation}
    We note that the time range of a single thread $I(\mu,\lambda,t)$ is $0$ s $\leq t\leq60$ s, so when $t<0$ s or $t>60$ s, we used the values at the boundary. Following \cite{2016Qiu}, $P(t)$ is set to be a two-half-Gaussian profile,
    \begin{equation}
    \begin{split}
     P(t) =
        \begin{cases} 
        \displaystyle C \, \mathrm{exp}\left[-\frac{(t-t_0)^2}{2\tau_0^2}\right] &  0< t < t_0 \\
        \displaystyle C \, \mathrm{exp}\left[-\frac{(t-t_0)^2}{18\tau_0^2}\right] &  t_0 <t < T \\
        \end{cases},
    \end{split}
    \label{FM_sum}
    \end{equation}
    with a quick growth and a long decay, where $C$ is the normalizing factor. Here, we take $T=1200$ s, $t_0=200$ s, and $\tau_0=50$ s; the normalizing factor $C$ is equal to $1/175\sqrt{2\pi}$. These parameters are set arbitrarily so that the shape of the soft X-ray flux curve in our model (Fig. \ref{rate}) is similar to that in a typical solar flare (e.g., \cite{2022Namekatac}). Other sets of $P(t)$ will definitely change the light curve, but they will not qualitatively change the results.
    As the heating of each thread lasts 60 s, the total simulation time would be 1260 s in the multi-thread model.

\subsection{Sun-as-a-star analysis}
\label{Sun-as-a-star analysis}

    As is well known, for an unresolved star we are actually observing its astrophysical flux $F$ (which is equivalent to the intensity averaged over the apparent stellar disk $\bar{I}$) instead of the specific intensity. For the quiet Sun ($t=0$ s in our simulation), the astrophysical flux is calculated as
    \begin{equation}
    \begin{split}
     \bar{I}(\lambda,0)&=2\int_{0}^{1} I(\mu,\lambda,0)\mu \mathrm{d}\mu\\     
     &=2\sum_{n=1}^{5}I(\mu_n,\lambda,0)\mu_n w_n,
    \end{split}
    \end{equation}
    where we used the Gauss-Legendre quadrature, and the weight $w_n$ is 0.118, 0.239, 0.284, 0.239, and 0.118 for the five $\mu_n$ values listed in Sect. \ref{Flare Modeling}.

    Compared to the whole disk, the flare is restricted to a small region where $\mu$ does not change much. We assume that the remainder of the disk is still well-represented by the quiet-Sun model. Then, for a given flare location $\mu$, we can write the difference in astrophysical flux as
    \begin{equation}
    \begin{split}
    \Delta \bar{I}(\mu,\lambda,t)&=\frac{1}{\pi R^2_\odot}\int_{A_\mathrm{flare}}\Delta I(\mu,\lambda,t)\mathrm{d}A\\
    &=\frac{\mu S_\mathrm{F}(I_F(\mu,\lambda,t)-I(\mu,\lambda,0))}{\pi R^2_\odot},
    \end{split}
    \label{deltas_equation}
    \end{equation}
    where the apparent area $A$ is given by $\mu S_\mathrm{F}$, and the flaring area $S_\mathrm{F}$ is assumed to be $1.5\times10^{18}\mathrm{cm}^2$, which is a typical value of the flare ribbon area (estimated from \cite{2022zhou}). After normalizing this quantity by the quiet-Sun flux at the far wing (near-continuum) of each line, we can obtain the so-called Sun-as-a-star spectrum $\Delta S$ as defined in \cite{2022Namekatac}:
    \begin{equation}
    \Delta S(\mu,\lambda,t)=\frac{\Delta \bar{I}(\mu,\lambda,t) }{\bar{I}(\lambda_\mathrm{cont},0)}.
    \end{equation}    
    We note that the astrophysical flux in a flare case is $\bar{I}(\mu,\lambda,t)=\bar{I}(\lambda,0)+\Delta \bar{I}(\mu,\lambda,t)$. After that, we constructed a grid of 50 flare models with ten different flare magnitudes and five different flare locations.
It should be pointed out that in the above method, the following assumptions have been made:
    \begin{enumerate}
        \item The rotation of the Sun is neglected. This might affect the Doppler shift of the Sun-as-a-star spectrum $\Delta S$ when the flare occurs near the east or west limb.
        \item The plane-parallel assumption means that we do not consider the orientation of the loops. The loop can be thought of as vertical to the surface, and the atmosphere is always extended in the $x$-$y$ plane. We did not consider whether the flaring atmosphere is embedded in a variety of different initial atmospheres, as suggested in previous observations \citep{2024Pietrowb}.
        \item The ``halo'' of the flare is not included in our 1D simulation, which can also play a role in contributing to the flare spectra \citep{2022Namekatac}. 
        \item The parameters of the nonthermal electron beam and the initial atmosphere are fixed in each multi-thread case. However, it is believed that beam injection sites can be localized within the flare ribbon \citep{2017Druett,2022Osborne,2023Polito}.
        \item We only focus on the flare ribbons as a simple model. Associated filament eruptions or CMEs are thus not included. However, as indicated by \cite{2022Otsu}, the Doppler shift of the filament can be a very prominent characteristic of the Sun-as-a-star spectrum.
        
    \end{enumerate}

\section{Sun-as-a-star flare spectra}
\label{Result}

    \subsection{Dependence on flare energy}
    \label{Dependence on flare energy}
    We first explore how the Sun-as-a-star flare spectra ($\Delta S$) are dependent on flare energy. Here, we fixed the flare location to around the disk center ($\mu$ = 0.953). 

\begin{figure*}[h]
    \centering
    \includegraphics[width=\textwidth]{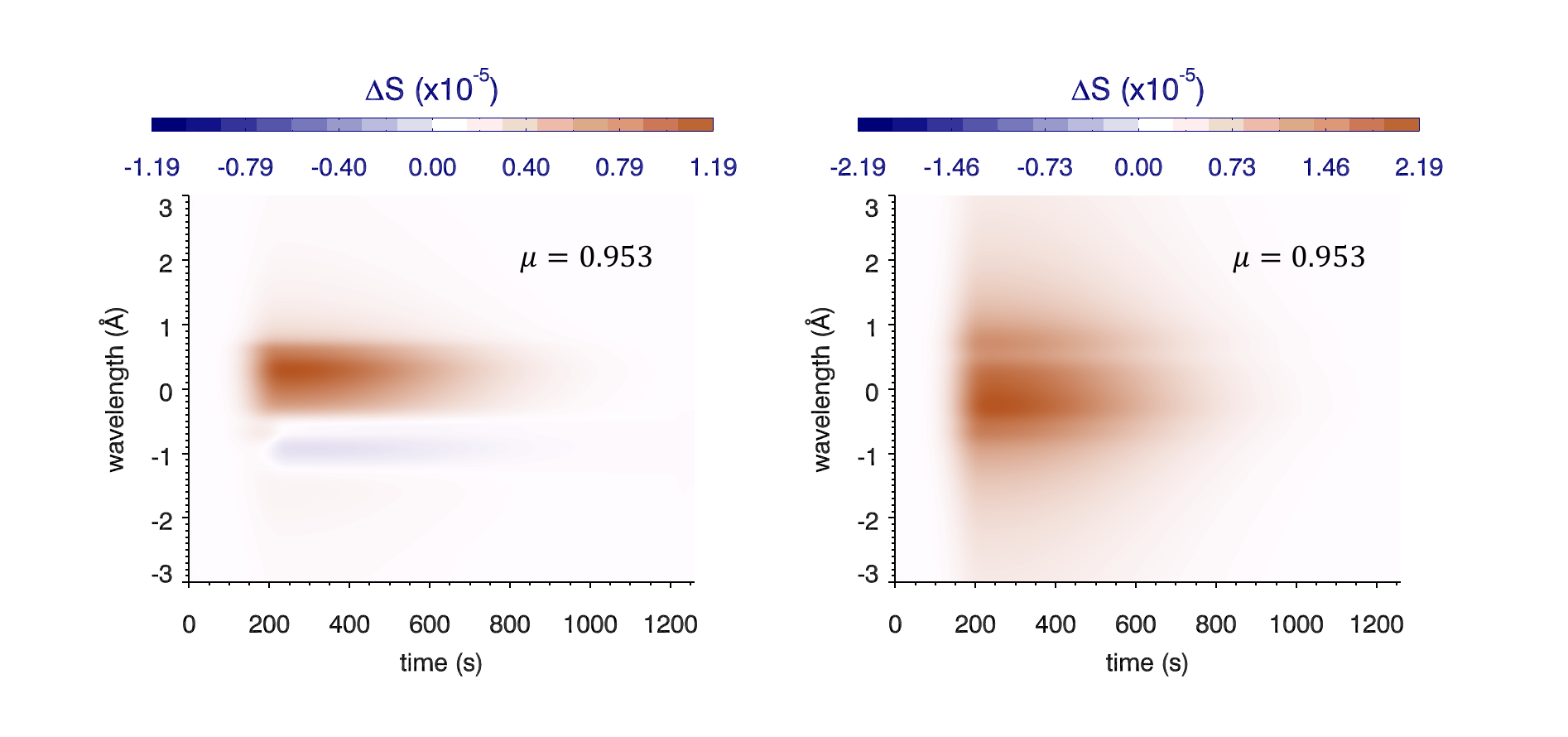}
    \caption{Time evolution of Sun-as-a-star spectrum $\Delta S$ of H$\alpha$ line with respect to wavelength and time for cases with $F_\mathrm{peak}=1$ (left) and $F_\mathrm{peak}=10$ (right) when flare occurs at the disk center ($\mu=0.953$). Red and blue represent enhancement and reduction, respectively.}
    \label{deltas}
\end{figure*}

    Figure \ref{deltas} shows the time evolution of $\Delta S$ of the H$\alpha$ line, in weak ($F_\mathrm{peak}=1$) and strong ($F_\mathrm{peak}=10$) flares. For the weak flare, after $t$ = 100 s the Sun-as-a-star spectrum within $\pm 1 \AA$ clearly starts to increase in intensity. At about $t$ = 200 s, $\Delta S$ in the blue wing becomes negative, while it is still positive in the red wing. Considering that the H$\alpha$ line is an absorption line, such an asymmetry is caused by the blueshift of the line \citep{2021Namekata}. For the strong flare, the profile is wider and stronger, and the value of $\Delta S$ is about twice that of the weak flare in the line center. Even in the far wing ($\pm 2 \AA$), $\Delta S$ shows a weak increase, which is caused by both the broadening of the line and the enhancement of the continua. In this case, the asymmetry caused by the line shift is not apparently seen, while the enhancement dominates and continues until the end. We checked the other eight cases between the weak and strong ones and find that the blueshift signal weakens as $F_\mathrm{peak}$ increases, which basically disappears at $F_\mathrm{peak}=5$.

    We now introduce three quantities that describe the properties of the line. The equivalent width $\Delta\mathrm{EW}$ is calculated simply by integrating $\Delta S$ over wavelength. The integration range is $\pm4 \AA$ here in order to include the major variations in the profile. The line asymmetry $A$ is obtained by the following equation: 
    \begin{center}
    \begin{equation}
    A(t) = \frac{\int^{\Delta\lambda}_0 \bar{I}(\lambda,t) \mathrm{d}\lambda - \int^0_{-\Delta\lambda} \bar{I}(\lambda,t) \mathrm{d}\lambda}{\int^{\Delta\lambda}_0 \bar{I}(\lambda,t) \mathrm{d}\lambda + \int^0_{-\Delta\lambda} \bar{I}(\lambda,t) \mathrm{d}\lambda}.
    \label{RB}
    \end{equation}
    \end{center}
    We followed Eq. 4 in \cite{2022Wu} and carried out an extra normalization. We set $\Delta\lambda$ to be $0.5 \AA$. This quantity $A(t)$ indicates red asymmetry for a positive value and blue asymmetry for a negative value. As for the line shift, we did not apply the two-component method (see \cite{2022Namekatac}) since $\Delta S$ shows irregular profiles rather than double peaks. Here, we chose the bisector method, which is generally used on solar flares \citep{1990Canfield}, that is, finding a wavelength $\lambda_m$ to make $\bar{I}(\lambda_m-\delta\lambda/2,t)=\bar{I}(\lambda_m+\delta\lambda/2,t)$, and here $\delta\lambda$ is set to 1 \AA. We chose to apply this method to $\bar{I}$ rather than $\Delta S$ because the shape of $\bar{I}$ is nearly symmetric after being averaged by the mainly quiet region of the disk. We can expect that the line shift values are quite small; thus, we primarily focused on the tendencies of these curves, which also contain useful information. 
   
\begin{figure*}[h]
    \centering
    \includegraphics[width=\textwidth]{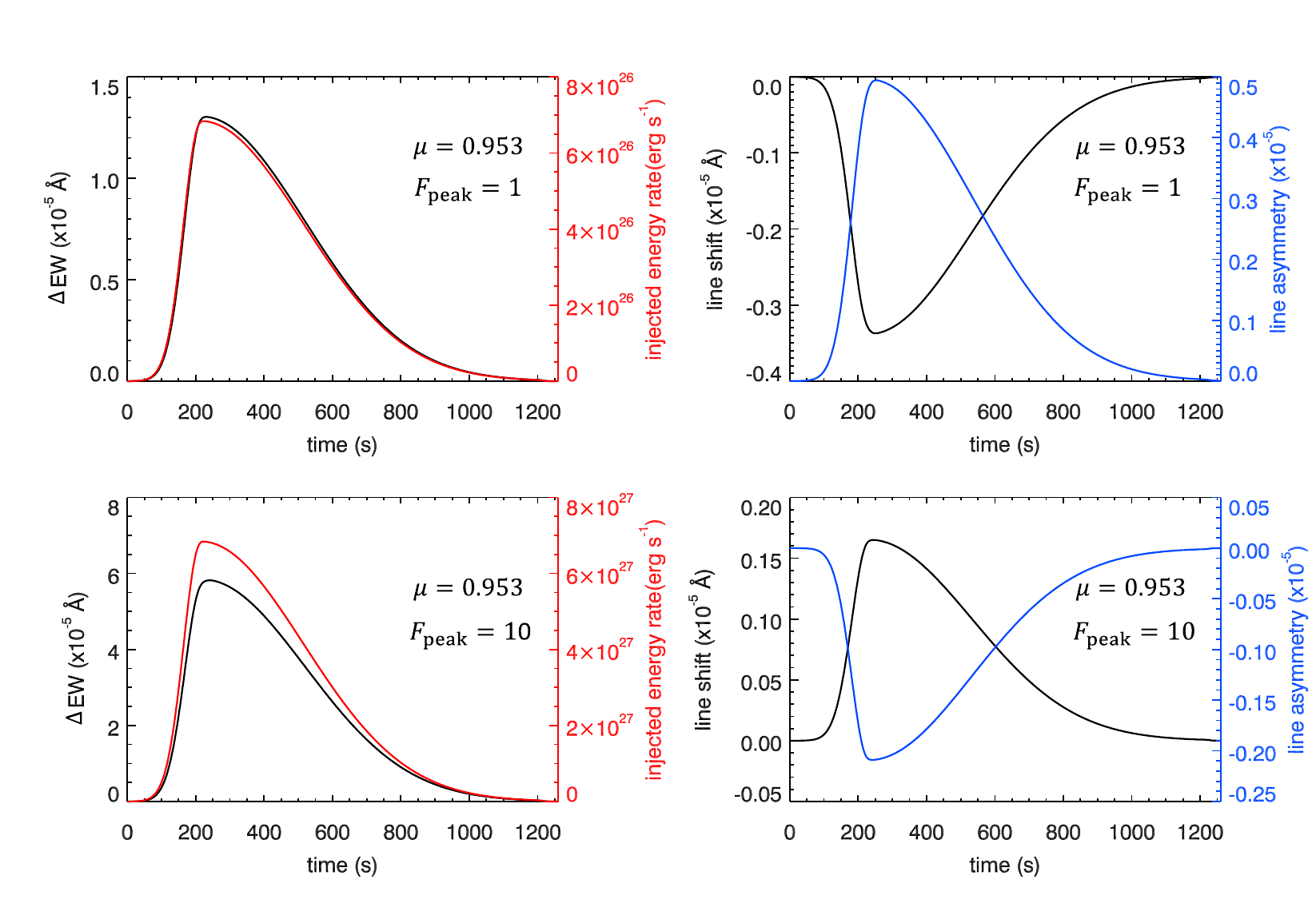}
    \caption{Quantities as functions of time. Left: Injected energy rate of nonthermal electron beam (red line) and equivalent width (black line). Right: Line shift measured from bisector method (black line) and red asymmetry calculated from Eq. \ref{RB} (blue line). Positive values represent redshift and red asymmetry.}
    \label{profile}
\end{figure*}

    The result is shown in Fig. \ref{profile}. For both weak and strong flares, the equivalent width increases rapidly and then decreases slowly, which follows the curve of the injected energy rate very well. The sign of the line shift and the line asymmetry depends on $F_\mathrm{peak}$: the weak flare presents a blueshift and red asymmetry, while the strong flare presents a redshift and blue asymmetry. After checking the other cases, we conclude that the line asymmetry of the H$\alpha$ line always shows a negative correlation with the line shift. 
    
\begin{figure*}[h]
    \centering
    \includegraphics[width=\textwidth]{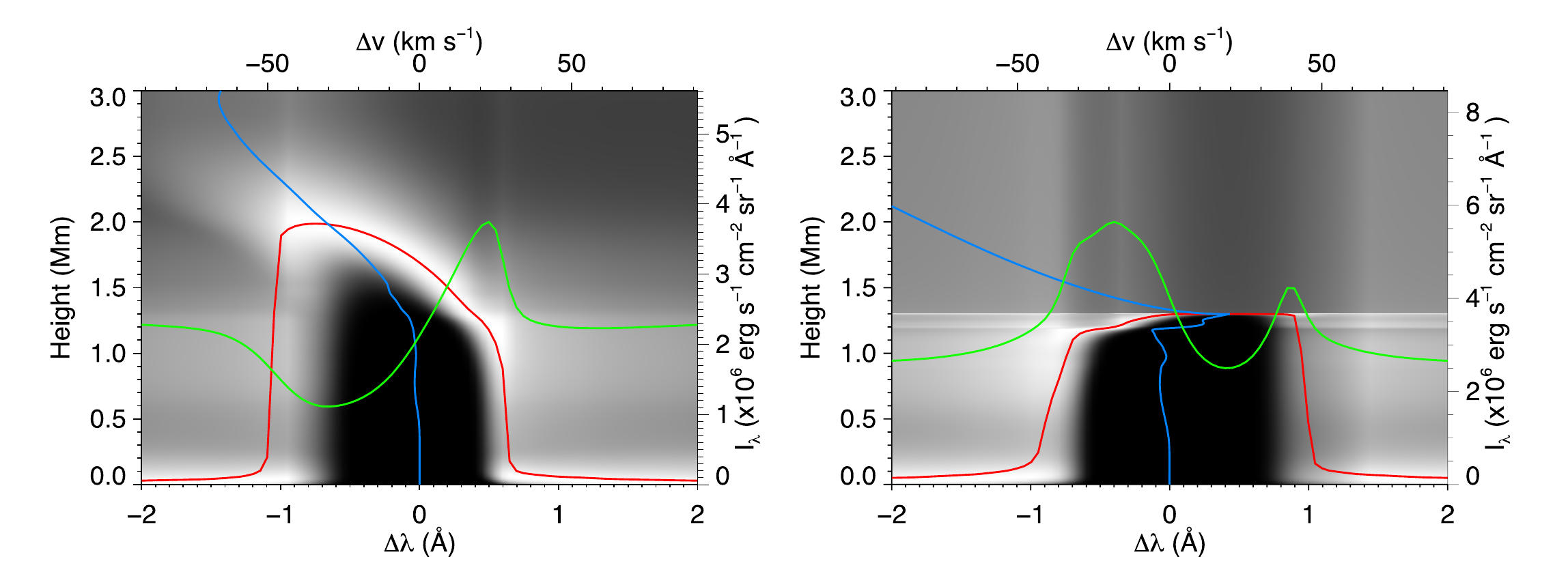}
    \caption{Line formation of H$\alpha$ line at $t=30$ s of single thread with $F_\mathrm{peak}=1$ (left) and $F_\mathrm{peak}=10$ (right). Lighter shades in the background denote larger values of the contribution function. The optical depth unity (red line) and the vertical velocity (blue line) as a function of height are overplotted, and the line profile is shown with a green line (right axis).}
    \label{contr}
\end{figure*}

    Focusing on the evolution tendency of the three quantities above, the curve of $\Delta\mathrm{EW}$ always follows that of the injected energy rate, but with a delay of tens of seconds. The curves of the line asymmetry and the line shift can have different behaviors. They might follow the energy rate curve as shown in the figure, and they might also deviate quite a lot as in cases $F_\mathrm{peak}=5$ and 6 (see Fig. \ref{appendix1}). Thus, we consider the $\Delta\mathrm{EW}$ of H$\alpha$ as a good ``real-time'' indicator of the flare rather than the other two, regardless of flare locations (refer to Fig. \ref{appendix2}). From the perspective of line formation, the $\Delta\mathrm{EW}$ depends highly on line enhancement and broadening. When the electron energy is released, and the line formation region is heated, the temperature there increases significantly, which immediately increases the $\Delta\mathrm{EW}$. However, the line asymmetry and the line shift result from the line-of-sight velocity of the plasma. Such a motion changes over simulation time \citep{2015Kuridze,2019Hong} and is not directly related to the injected energy. It is worth mentioning that \cite{2022Namekatac} proposed the AIA 1600 \AA\ curve as a proxy of the time profile of the energy release, which is somehow similar to the equivalent width.
    
    As $F_\mathrm{peak}$ increases, the blueshift as measured from the Sun-as-a-star spectrum changes to the redshift. To test whether the measured quantity reflects real motions in the atmosphere, we calculated the contribution function of $t=30$ s of a single thread in both cases (see Fig. \ref{contr}). The contribution function is defined as $C_{\lambda}(z)=\frac{1}{\mu}j_{\lambda}(z)e^{-\tau_{\lambda}(z)/\mu}$, where $j_{\lambda}$ and $\tau_{\lambda}$ are the emissivity and optical depth, respectively. Integrating $C_{\lambda}$ along the height, we obtain the emergent intensity. When $F_\mathrm{peak}=1$, the line forms in the upper chromosphere (about 1.5Mm). In this region, the plasma moves upward, which corresponds to the blueshift. As for case $F_\mathrm{peak}=10$, the line forms lower and in a more concentrated way (about 1.3Mm); the plasma moves downward and shows the redshift in our result. We checked the other eight cases and find that the plasma in the line formation region gradually shifts from upward movement to downward movement with increasing $F_\mathrm{peak}$, with the turning point at around $F_\mathrm{peak}=6$. This accounts for the weakened blueshift signal as $F_\mathrm{peak}$ increases. To account for the relationship between the line asymmetry and the line shift, we used the same interpretation as \cite{2015Kuridze}: upward motion in the line formation region shifts the line core to the blue side, meaning more photons are absorbed in the blue wing. Consequently, the intensity in the red wing is larger compared to its blue wing counterpart, resulting in a red asymmetry. A similar explanation is used for redshift and blue asymmetry. The picture of $C_{\lambda}$ shows that the bisector method is reasonable for detecting plasma motion in the chromosphere, at least for the velocity sign.
        
    \subsection{Dependence on flare location}
    \label{angle dependent}
    According to the Eddington-Barbier approximation -- $I_{\nu}(\mu) \approx S_{\nu}(\tau=\mu)$ -- the emergent intensity of location $\mu$ depends on the source function where $\tau = \mu$. The line intensities at different locations reflect different layers of the solar atmosphere. Closer to the limb, $\mu$ is smaller; thus, we have properties in the shallower atmosphere, which causes the center-to-limb variation. In brief, the spectra show different profiles with respect to location, whether in quiet or flare regions.

    \begin{figure*}[h]
    \centering
    \includegraphics[width=\textwidth]{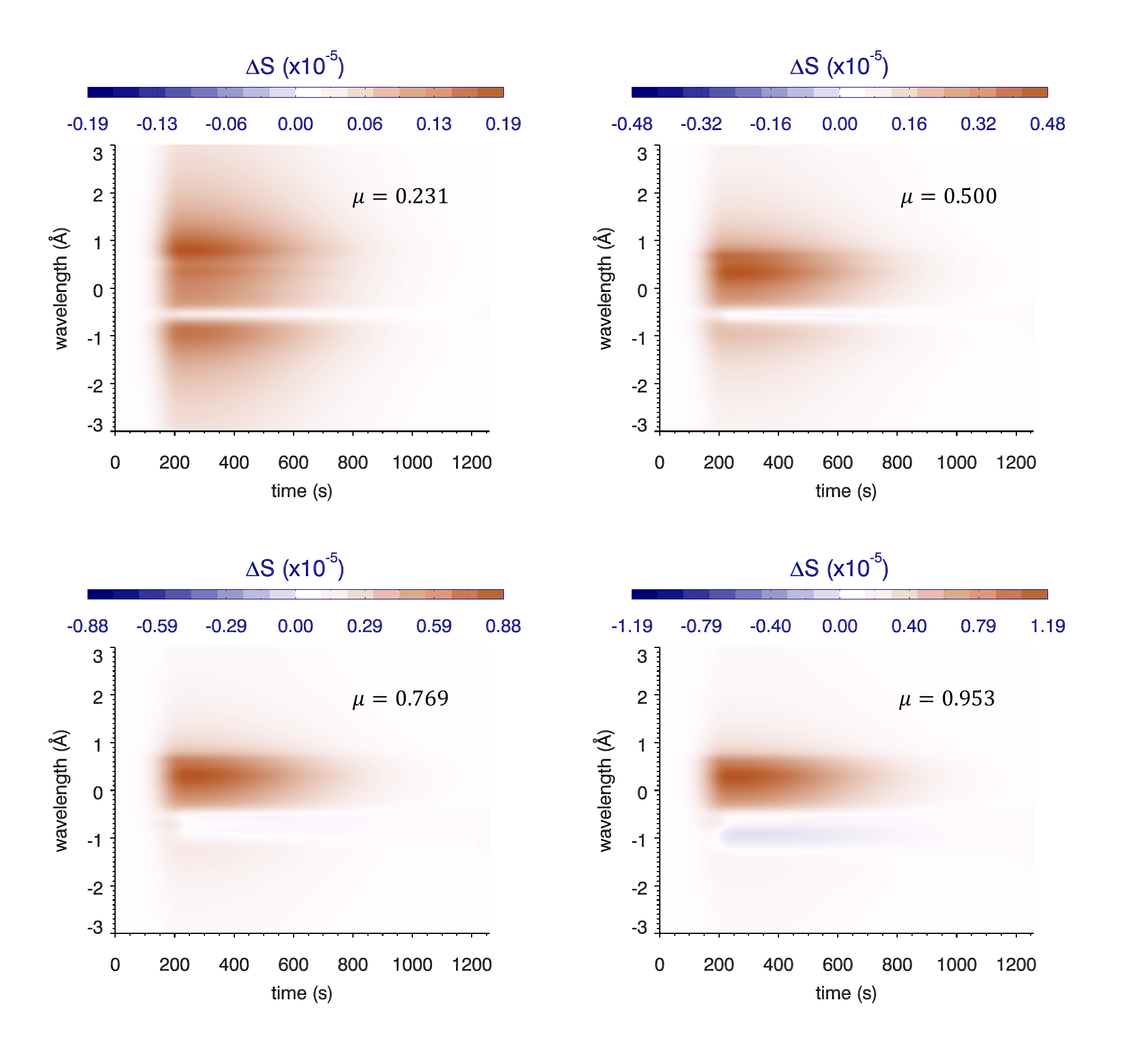}
    \caption{Sun-as-a-star spectrum $\Delta S$ of H$\alpha$ line of wavelength and time at different flare location $\mu$ when $F_\mathrm{peak}=1$. The rest are the same as in Fig. \ref{deltas}.}
    \label{angle_loop}
    \end{figure*}

    \begin{figure*}[h]
    \centering
    \includegraphics[width=\textwidth]{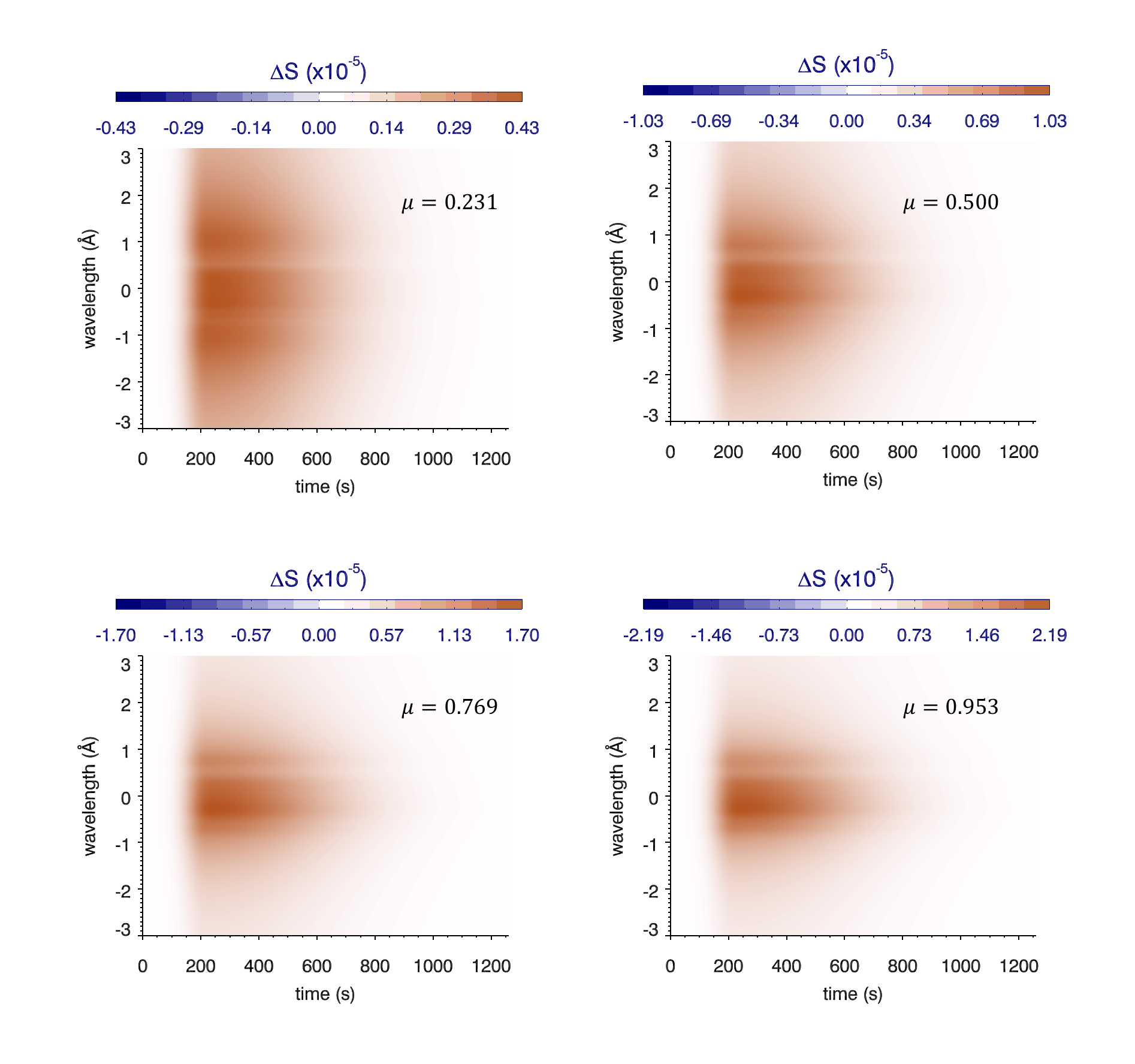}
    \caption{Same as Fig. \ref{angle_loop}, but $F_\mathrm{peak}=10$.}
    \label{angle_loop2}
    \end{figure*}
    
    Figure \ref{angle_loop} presents the variation of $\Delta S$ at different locations when $F_\mathrm{peak}=1$. Case $\mu=0.047$ is not shown, because the $\Delta S$ of such an event is too weak to be detected after being multiplied by $\mu$ (Eq. \ref{deltas_equation}). At each $\mu$, the spectra show apparent red asymmetry and line broadening, which are similar to those found by \cite{2022Namekatac}. From $\mu=0.953$ to 0.231, the maximum value of $\Delta S$ reduces because the projection area of the flare becomes smaller. We then focused on the profiles of $\Delta S$ when $\mu$ decreases. On one hand, the enhancement is more concentrated in the line core when $\mu=0.953$, but it has a larger spread when $\mu=0.231$. On the other hand, the profile shows multi-peaks, with most lying in the line wing instead of the line center. The absorption dip in the blue wing has negative values at the disk center, but it is less pronounced toward the solar limb. For strong flares (Fig. \ref{angle_loop2}), the variations in different locations are similar to the weak ones, but with generally larger values of $\Delta S$ and a redshifted absorption dip. 
    
\begin{figure*}[h]
    \centering
    \includegraphics[width=\textwidth]{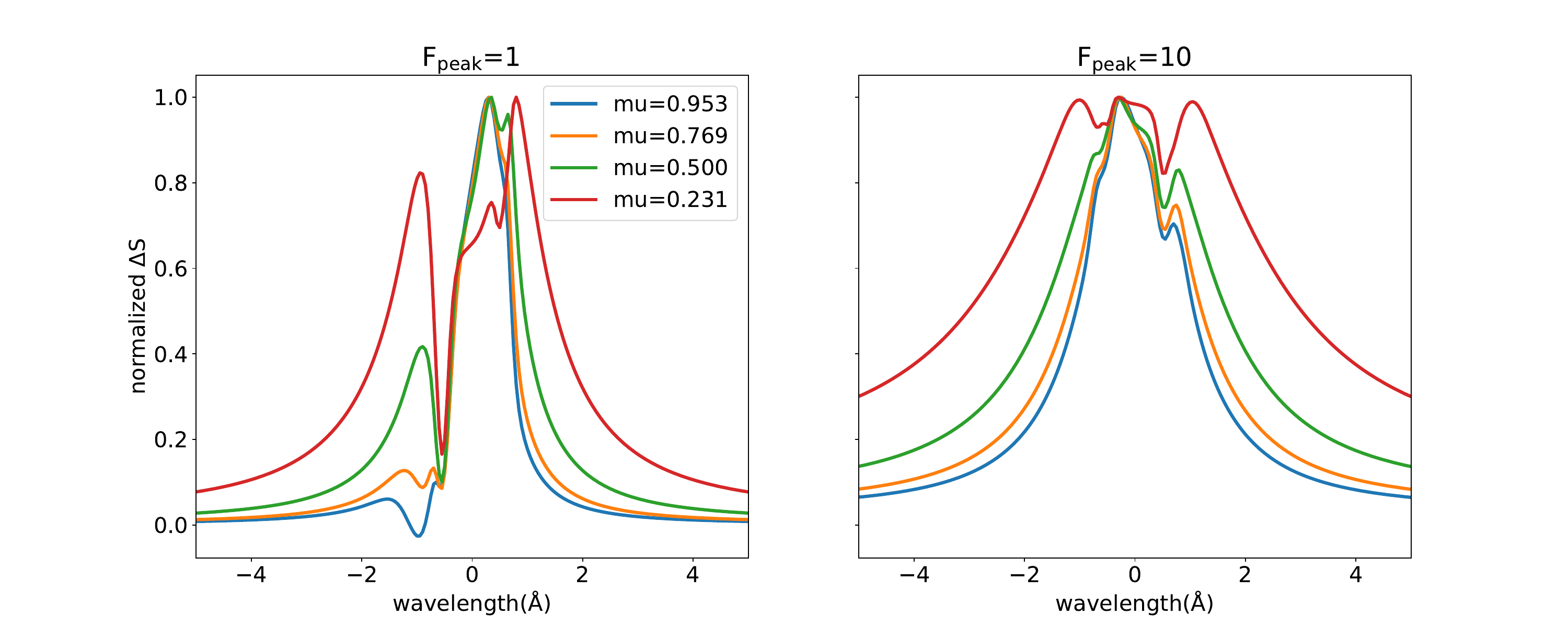}
    \caption{Normalized profile $\Delta \Hat{S}$ of   H$\alpha$ line at different locations $\mu$ when $F_\mathrm{peak}=1$ (left) and $F_\mathrm{peak}=10$ (right). The spectrum of the limb flare is relatively wider and shows a dip in the line core.}
    \label{normalizedS}
    \end{figure*}

    In order to focus more on profile shapes rather than values, we normalized $\Delta S$ by the maximum value of the whole profile at each time: $\Delta \Hat{S}(\lambda,\mu,t)=\Delta S(\lambda,\mu,t)/\mathrm{max}(\Delta S(\lambda,\mu,t))$. Then, we plotted $\Delta \Hat{S}$ over different $\mu$ in cases $F_\mathrm{peak}=1$ and 10 in Fig. \ref{normalizedS}. We show the result of $t$ = 200 s, which is the peak time of our flare model. In the weak flare, values of $\Delta \Hat{S}$ at the red wings are larger than those at the blue wings, which is mainly caused by the blueshift, as discussed in Sect. \ref{Dependence on flare energy}. For the strong flare, the result is exactly the opposite: redshift decreases the values at the red wings and increases them at the blue wings. 

    As noted above, the width and dip of the normalized profile $\Delta \Hat{S}$ vary with $\mu$ and $F_\mathrm{peak}$. For a quantitative analysis, we define the following two quantities to extract useful information from the profiles.
One is the line width, defined as the second moment of the normalized profile $\Delta \Hat{S}$,
    \begin{center}
    \begin{equation}
    W=\sqrt{\frac{\int^{\Delta\lambda}_{-\Delta\lambda}(\lambda-\lambda_0)^2\Delta\Hat{S}\mathrm{d}\lambda}{\int^{\Delta\lambda}_{-\Delta\lambda}\Delta\Hat{S}\mathrm{d}\lambda}},
    \end{equation}
    \end{center}
    where $\lambda_0$ is the wavelength at the line center, and the value of $\Delta\lambda$ is set to $4\AA$. 
    
\begin{figure}[h]
    \centering
    \includegraphics[scale=0.22]{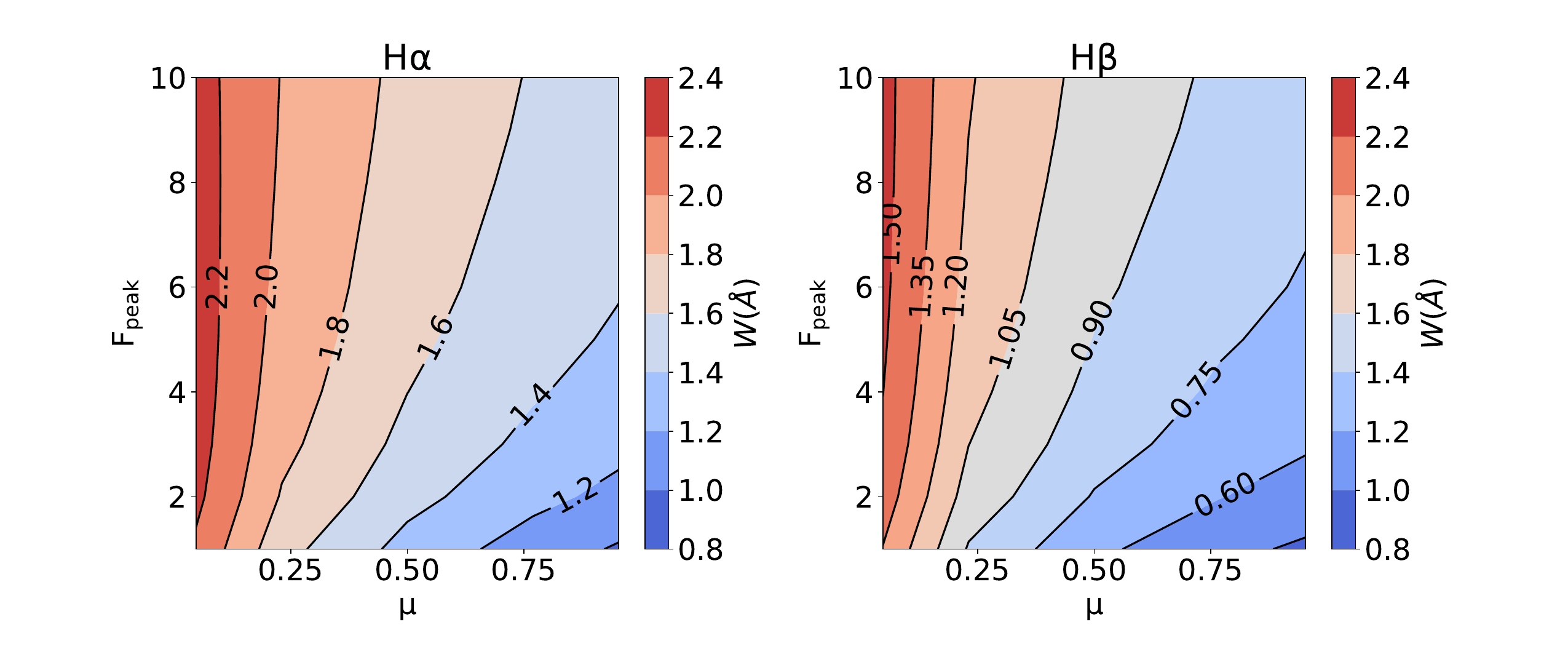}
    \caption{Filled contours of line width $W$ of H$\alpha$ (left) and H$\beta$ (right) line as a function of $\mu$ and $F_\mathrm{peak}$.}
    \label{LW}
    \end{figure}

    We calculated the line width of both H$\alpha$ and H$\beta$ lines in different models and show the results in Fig. \ref{LW}. The patterns in these two lines are quite similar. A larger $F_\mathrm{peak}$ and a smaller $\mu$ would result in a larger $W$, which is also revealed from Fig. \ref{normalizedS}. The value of $W$(H$\alpha$) varies from 1.0 (the smallest $F_\mathrm{peak}$ and the largest $\mu$) to 2.3 (the largest $F_\mathrm{peak}$ and the smallest $\mu$), while the value of $W$(H$\beta$) is smaller, ranging from 0.4 to 1.5. We also note that the dependence on $\mu$ is more pronounced than that on $F_\mathrm{peak}$. For smaller values of $\mu$, the line width does not vary much with $F_\mathrm{peak}$.
    
    The other quantity is d$\Delta S$, which measures the relative difference between the line core and line wings:
    \begin{center}
    \begin{equation}
    \mathrm{d}\Delta S=\frac{\int_{\lambda_C}\Delta \Hat{S}\mathrm{d}\lambda-0.5(\int_{\lambda_R}\Delta \Hat{S}\mathrm{d}\lambda+\int_{\lambda_B}\Delta \Hat{S}\mathrm{d}\lambda)}{\int_{\lambda_C}\Delta \Hat{S}\mathrm{d}\lambda}, 
    \end{equation}
    \end{center}    
    where $\lambda_C$, $\lambda_R$ and $\lambda_B$ are integration ranges for the line core, red wing, and blue wing. We chose $\lambda_C$ = [-0.5 \AA , 0.5 \AA], $\lambda_R$ = [0.5 \AA , 1.5 \AA] and $\lambda_B$ = [-1.5 \AA , -0.5 \AA] for the H$\alpha$ line. For H$\beta$, the ranges of $\lambda_C$, $\lambda_R$, and $\lambda_B$ are set to [-0.22 \AA, 0.22 \AA], [0.22 \AA, 0.66 \AA], and [-0.66 \AA, -0.22 \AA], respectively. A negative d$\Delta S$ indicates a dip in the line core. 

\begin{figure}[h]
    \centering
    \includegraphics[scale=0.22]{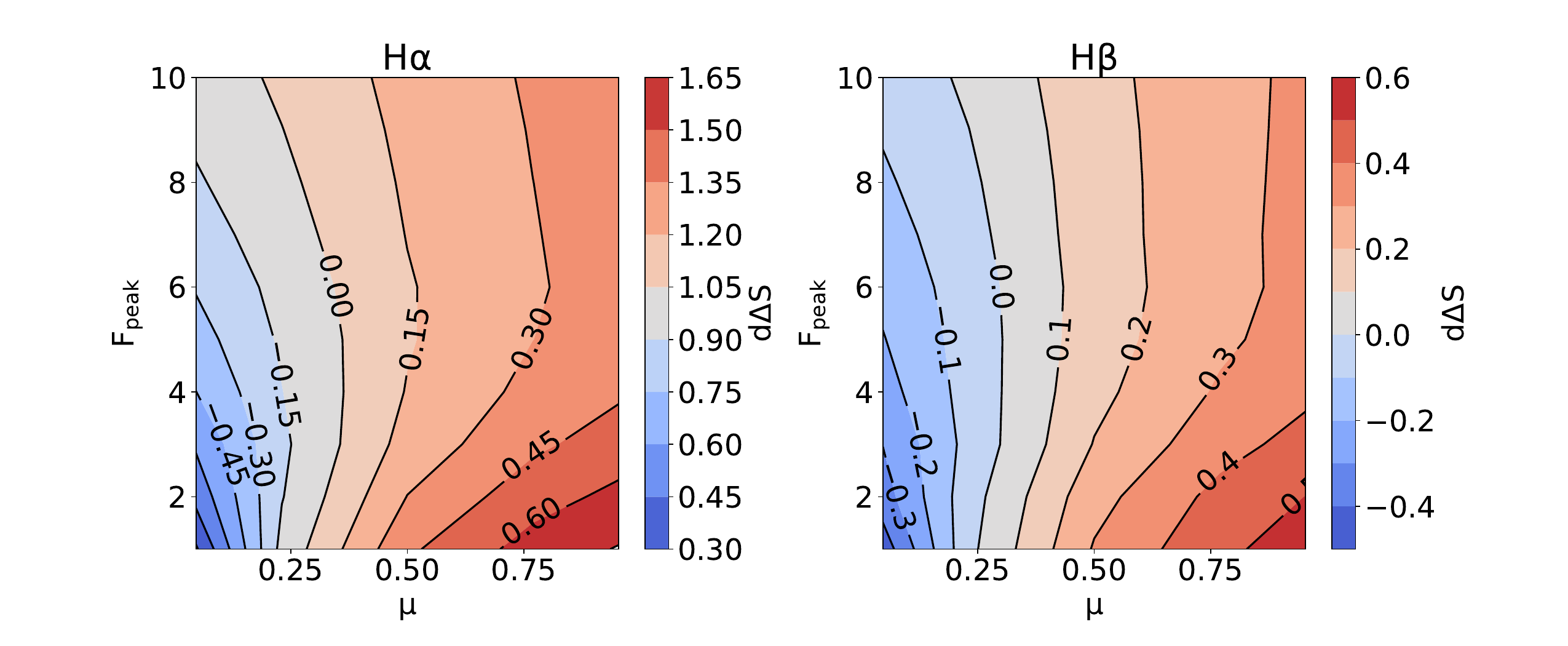}
    \caption{Same as Fig. \ref{LW}, but for d$\Delta S$.}
    \label{dds}
    \end{figure}

    The results of d$\Delta S$ are plotted in Fig. \ref{dds}. For both lines, the value of d$\Delta S$ has a weak dependence on $F_\mathrm{peak}$ for $0.2 < \mu <0.4$. The dependence on $F_\mathrm{peak}$ becomes more pronounced when $\mu$ is larger than 0.6 or smaller than 0.2. For both $W$ and d$\Delta S$, the dependence on $\mu$ is always strong regardless of the value of $F_\mathrm{peak}$. These two contours can be combined as diagnostics of the flare location from the observed line profiles, which are discussed further in Sect. \ref{Determination of the flare location from observations}.

\section{Comparison with observations}
    \label{comparing to observation}
    \subsection{Contrast profiles}
    \label{activity indices}
\begin{figure*}[h]
    \centering
    \includegraphics[width=\textwidth]{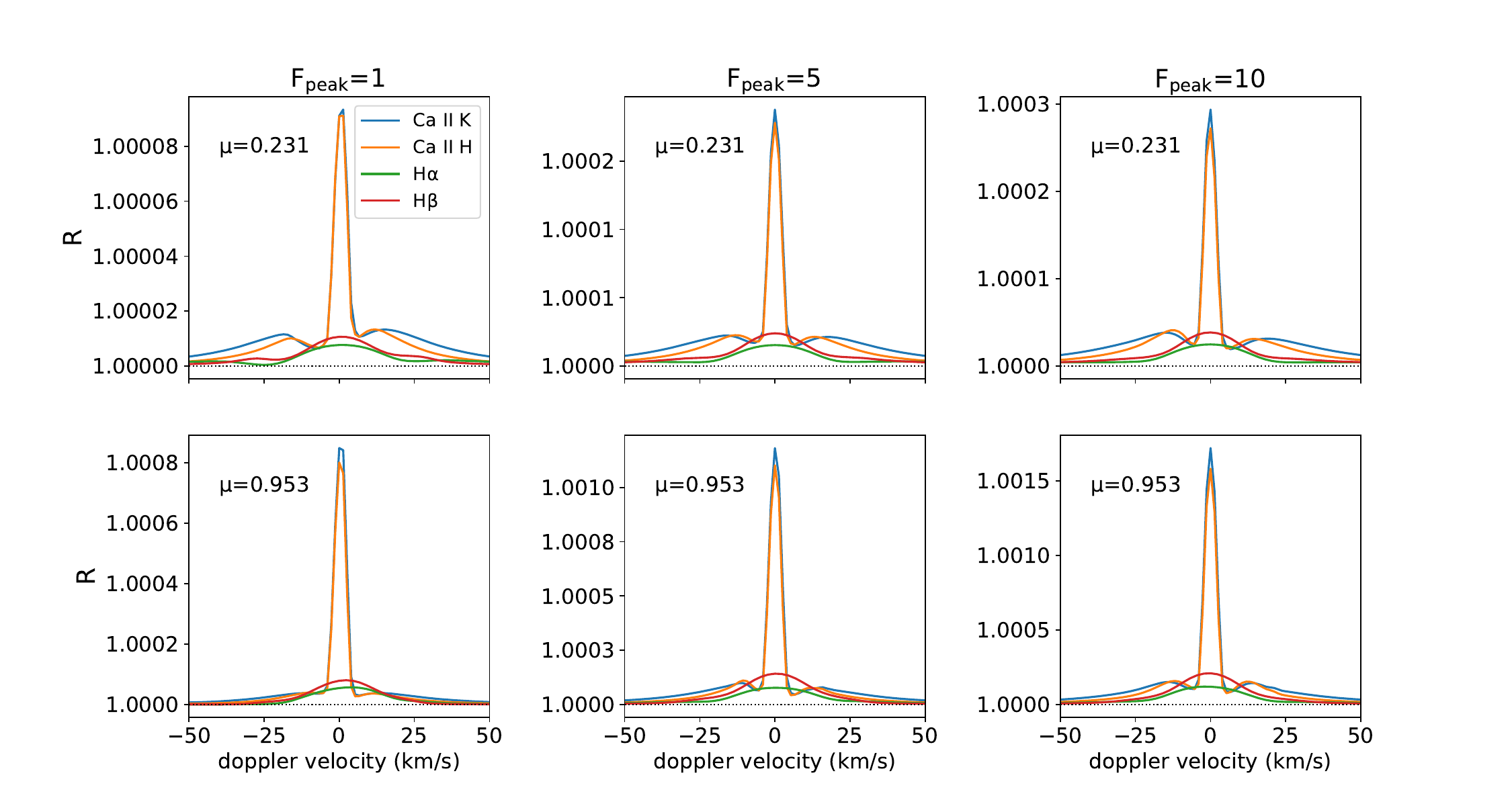}
    \caption{Contrast profile $R$ of H$\alpha$, H$\beta$, \ion{Ca}{II} K, and \ion{Ca}{II} H lines in different models. The parameter $F_\mathrm{peak}$ varies among 1 (Left), 5 (middle), and 10 (right), while $\mu$ varies between 0.231 (top) and 0.953 (bottom). }
    \label{contrast profile}
    \end{figure*}

    Aside from the method mentioned in Sect. \ref{Sun-as-a-star analysis}, the other approach to analyzing the Sun-as-a-star spectrum is to calculate the contrast profile \citep{2024Pietrow}. The contrast profile $R$ is obtained by normalizing $\bar{I}(\mu,\lambda,t)$ with the quiet spectrum: $R(\mu,\lambda,t)=\bar{I}(\mu,\lambda,t)/\bar{I}(\mu,\lambda,0)$. When normalizing, the values of the quiet spectrum on the denominator vary over wavelength, rather than a constant as in Eq. \ref{deltas_equation}, so shapes of $R$ evidently differ from $\Delta S.$ We show $R$ of four synthesized chromospheric lines (H$\alpha$, H$\beta$, \ion{Ca}{II} K, and \ion{Ca}{II} H) with respect to two parameters at $t$ = 200 s in Fig. \ref{contrast profile}. The shapes of the H$\alpha$ and H$\beta$ contrast profiles look similar, which is also the case between the \ion{Ca}{II} K and \ion{Ca}{II} H profiles. However, the H$\beta$ and \ion{Ca}{II} K contrast profiles are wider and stronger than those of  H$\alpha$ and \ion{Ca}{II} H, respectively. In these cases, the contrast profiles of \ion{Ca}{II} are stronger and wider than that of \ion{H}{I}, which indicates that \ion{Ca}{II} lines are more sensitive to flares under nonthermal electron beam heating. As $F_\mathrm{peak}$ increases and $\mu$ decreases, the triple-peak profiles of the \ion{Ca}{II} lines become more pronounced. 
    
    Compared to observations of the X-class flares in \citep{2024Pietrow}, the intensity of $R$ in our simulation is much smaller, even if for the strongest case ($F_\mathrm{peak}$=10, $\mu$=0.953). The observed contrast profiles of each line (particularly \ion{Ca}{II} lines) are also wider than simulations. One reason for these differences is that our flare area is not as large as those in their observations. We used the same flare area ($1.5\times10^{18}\mathrm{cm}^2$) to calculate the soft X-ray flux in our models following \cite{2015Pinto}, assuming the radiation mechanism to be a thermal bremsstrahlung one. An optically thin assumption is also employed here, meaning that the X-ray flux is independent of $\mu$. The corresponding GOES soft X-ray flux is partly shown in Fig. \ref{index}. We find that for different flares in our simulation, as the electron beam flux increases, the increase in soft X-ray flux is larger than the increase in the contrast profiles. This implies that in order to reach the same soft X-ray flux level, a larger flare area with a smaller beam flux would result in a larger value in the contrast profile.  \cite{2024Pietrow} also noted a red wing enhancement in \ion{Ca}{II} K of the observed X9.3 flare implying chromospheric condensation (Malcolm Druett, private communication), while the red wing enhancement only appears in both weak and limb flares in our models (see the upper left panel of Fig. \ref{index}), and is the result of weak-upflowing absorptive plasma at the bottom of the evaporation region (see the left panel of Fig. \ref{contr}).
    
\subsection{Activity indices}
    Several activity indices of chromospheric lines have been defined to quantify the strength of stellar chromospheric activities \citep{2009Boisse,2020Melbourne,2023Amazo-Gomez}. In the quiet-Sun profiles, the line centers of these lines form in the chromosphere, while the line wings form in the photosphere. During flaring activities, especially when heating is dominated by a nonthermal beam, the upper chromosphere is heated more efficiently than the lower layers. Consequently, the line center that is formed in the chromosphere exhibits a stronger response than the wings and the continuum. This result holds even though flares can produce additional strong chromospheric sources in the line wings and bound-free continua \citep{2015Allred,2017Simoes,2018Druett}. \cite{1978Vaughan} introduced the concept of S-index, which is the ratio between the line core of \ion{Ca}{II} H/K and the nearby continuum. 
    
    In quiet-Sun profiles, the line centers form in the chromosphere, while the line wings form in the photosphere. During flaring activities, especially when heating is dominated by a nonthermal beam, the upper chromosphere is heated more efficiently than the lower layers. Consequently, the line center that is formed in the chromosphere exhibits a stronger response than the wings and the continuum. We followed \cite{2020Melbourne} to calculate the S-index as 

    \begin{equation}
    \alpha_{\mathrm{H/K}}=\frac{F_\mathrm{H}+F_\mathrm{K}}{F_\mathrm{R}+F_\mathrm{V}},
    \end{equation}
    where the line fluxes $F_\mathrm{H}$ and $F_\mathrm{K}$ are the integration of the astrophysical flux over the 1.09 \AA\ wide triangular bands centered on 3933.66 \AA\ and 3968.46 \AA , and the near-continuum fluxes $F_\mathrm{R}$ and $F_\mathrm{V}$ are integrated over $\pm10$ \AA\ centered on 3901 \AA\ and 4001 \AA, respectively. The H$\alpha$-index is calculated as
    \begin{equation}
    \alpha_\mathrm{H\alpha}=\frac{F_\mathrm{H\alpha}}{F_1+F_2},
    \end{equation}
    where $F_\mathrm{H\alpha}$ is the integrated flux over 0.6 \AA\ around the H$\alpha$ line center (6562.8 \AA) and $F_1$ and $F_2$ correspond to the continuum flux integrated over 8 \AA\ windows centered at 6550.85 and 6580.28 \AA , respectively. Similarly, for H$\beta$, the index is obtained by setting the wavelength centers of the integration ranges of $F_\mathrm{H\beta}$, $F_1,$ and $F_2$ to 4861.38, 4585, and 4880 \AA, with a window width of 0.2, 10, and 10 \AA, respectively. Wavelength window values are the same as those of \cite{2024Pietrow}, who use relatively narrower windows in the line center to reduce the wash-out effect by the photosphere. \cite{2024Pietrow} suggests that the H$\alpha$ and H$\beta$ index values are strongly dependent on the width of these windows, so here we only compare our results to their observations.

\begin{figure*}[h]
    \centering
    \includegraphics[width=\textwidth]{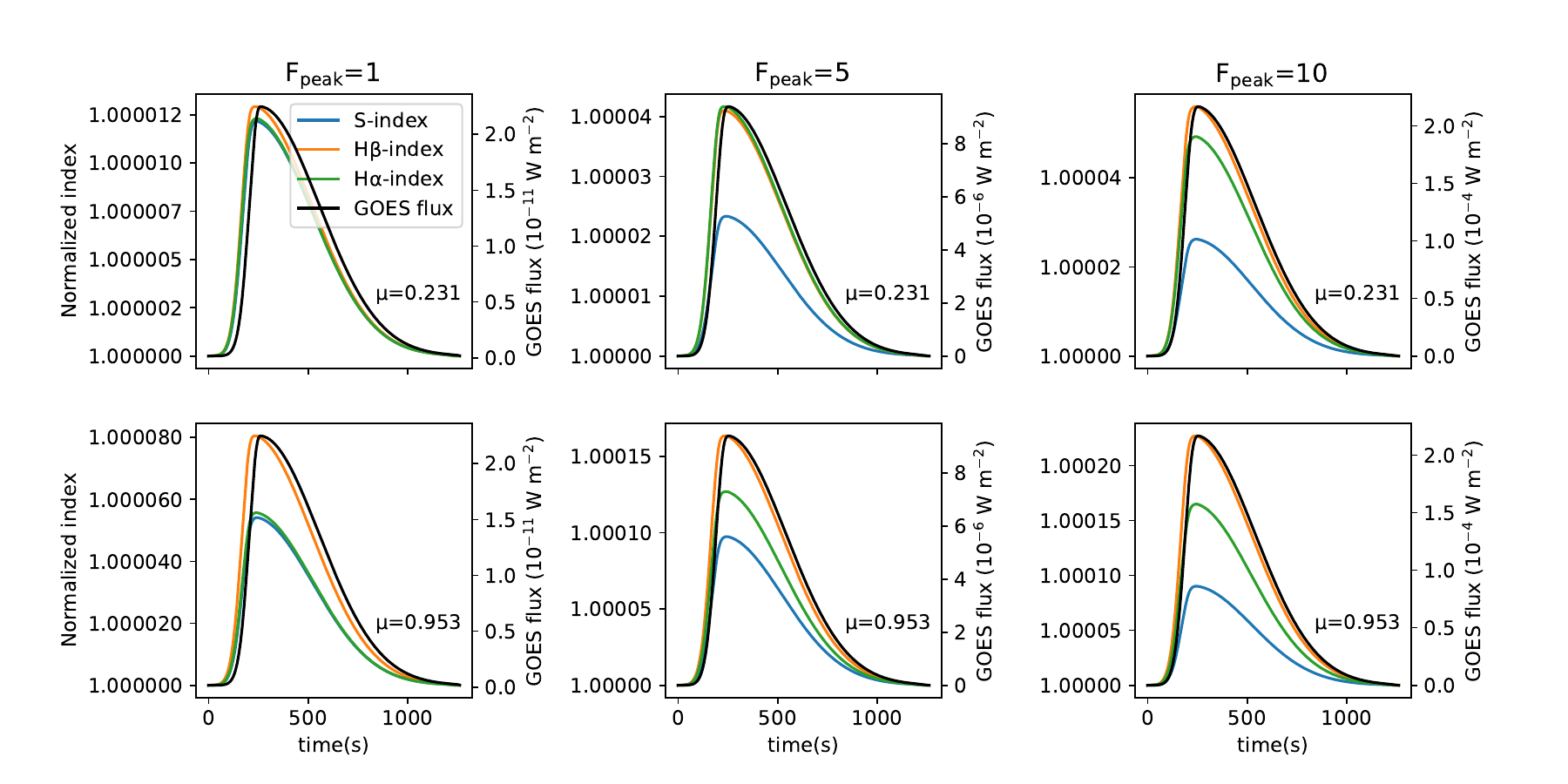}
    \caption{Time evolution of S-index, H$\alpha$-index, and H$\beta$-index; arrangement order of parameters is similar to that in Fig. \ref{contrast profile}. }
    \label{index}
    \end{figure*}

    We plot the time evolution of the three indices in Fig. \ref{index}. The indices in the observation often exhibit much more noise compared to our simulation, which explains why only strong flares are detectable from the indices in the observations. Considering the X9.3 flare reported by \cite{2024Pietrow}, it results in an approximately 1\% increase of the S-index and a 0.5\% increase of the H$\alpha$/H$\beta$ index. In our simulation, even for the strongest case (bottom right panel), we see an approximately 0.02\% increase of the H$\alpha$/H$\beta$-index and a 0.01\% increase of the S-index, which is nearly one fifth of that in \cite{2024Pietrow}. As discussed above, one possible reason is that the flare class is not large enough. Another reason is that the contrast profiles are narrower than those in \cite{2024Pietrow}. For case $F_\mathrm{peak}=10$, $\mu=0.953$ (bottom right panel in Fig. \ref{contrast profile}), the double peaks of the \ion{Ca}{II} K line wings are located at $\pm$ 10 km/s, but at about $\pm$ 70 km/s in the observation. The narrow profiles result in a much smaller increase of $F_\mathrm{H}$ and $F_\mathrm{K}$, leading to a much smaller increase of the S-index. Considering the noise level in \cite{2024Pietrow}, it could still be difficult to detect the flare signal from the activity indices of our simulated X2 flare. \cite{2024Pietrow} also reveals that the increase of the S-index has a longer duration compared to the other two indices. However, the longer duration of the S-index is not found in our simulation, since our simulation is only in one-dimensional, and we did not include the complexity of the multidimensional evolution and heating in solar flares \citep{2024druett,2024Ruan}.
    
    In solar flare observations, the soft X-ray flux of GOES 1--8 {\AA} is also considered as an index to indicate flare magnitudes. We show the flux together with the three indices in Fig. \ref{index}. The GOES flux for the case of $F_\mathrm{peak}=1$ is notably smaller due to the neglect of coronal heating by electrons below the cut-off energy. The peak of the soft X-ray curve has an approximately 30 s time delay compared to other index curves. Such a time delay is similar to those of observations reported in \cite{2022Namekatac}, where the $\Delta\mathrm{EW}$ of the H$\alpha$ peaks faster than GOES X-ray. It seems that the heating of the chromosphere is faster than that of the corona in our model, which is a commonly observed relationship in solar flares and is one aspect of the relationship between the gradient of the soft X-ray flux curve and the hard X-ray flux curve known as the Neupert effect \citep{1968Neupert}.

    \subsection{Diagnosing the flare location from observations}
    \label{Determination of the flare location from observations}
    One of the primary objectives of our work is to find a method to diagnose the location of a stellar flare. In Sect. \ref{angle dependent}, we introduced two quantities -- $W$ and d$\Delta S$ -- and plotted their variations with respect to two parameters, $\mu$ and $F_\mathrm{peak}$. Both $W$ and d$\Delta S$ have a larger dependence on $\mu$, and these contours can be used to locate the flare. For example, if d$\Delta S($H$\alpha)<0$, then $\mu$ should be less than 0.4 from Fig. \ref{dds}, which makes it a near-limb event. By combining Figs. \ref{LW} and \ref{dds}, one can determine the possible range of $\mu$ and $F_\mathrm{peak}$. Taking the H$\beta$ line as an example, if $W=0.6$ and d$\Delta S=0.4$, we can diagnose that $F_\mathrm{peak}$ is small and $\mu$ is large, meaning that a weak flare occurs near the disk center.

    We point out that this diagnostic method only works well when the flare ribbon causes the majority of the variation of the Sun-as-a-star spectrum. In real observations, solar flares are sometimes accompanied by other transient phenomena such as surges, filament eruptions, and coronal rain. Usually, these events can be identified by their unique characteristics. For example, a sudden enhancement of $\Delta S$ shifting from the blue wing to the red wing may indicate an out-of-limb prominence eruption \cite{2022Otsu}. An absorption of $\Delta S$ shifting from the blue wing to the red wing indicates a filament eruption or a surge on the solar disk \citep{2024Ikuta}. Nonetheless, as long as we exclude the interference events above, the characteristics of the spectrum can be systematically compared with our simulation. We take event 2 and event 4 of \cite{2022Otsu} as examples. The Sun-as-a-star spectrum of the limb flare (event 2) seems to appear as a central dip (see their Fig. 4), while that of the non-limb flare (event 4) shows a single-peak profile (see their Fig. 8), similarly to our results.
        
    \subsection{From flux to energy}
    \label{unit conversion}
    As mentioned in Sect. \ref{method}, the astrophysical flux $F$ (equivalent to $\bar{I}$) is measured in units of $\mathrm{erg}\ \mathrm{s^{-1}}\ \mathrm{cm^{-2}}\ \mathrm{sr^{-2}}\ \mathrm{\AA^{-1}}$. In order to be compared with the Sun-as-a-star observation, $\bar{I}$ must be converted into flux received by the detector (also known as the irradiance if the detector is on the Earth) as follows:
\begin{equation}
    \label{assumption}
    \mathcal{R}(\mu,\lambda,t)=\frac{\pi R_\odot^2}{D^2}\bar{I}(\mu,\lambda,t),
\end{equation}
    with the solar radius $R_\odot$ and distance $D$. 
    
\begin{figure*}[h]
    \centering
    \includegraphics[width=\textwidth]{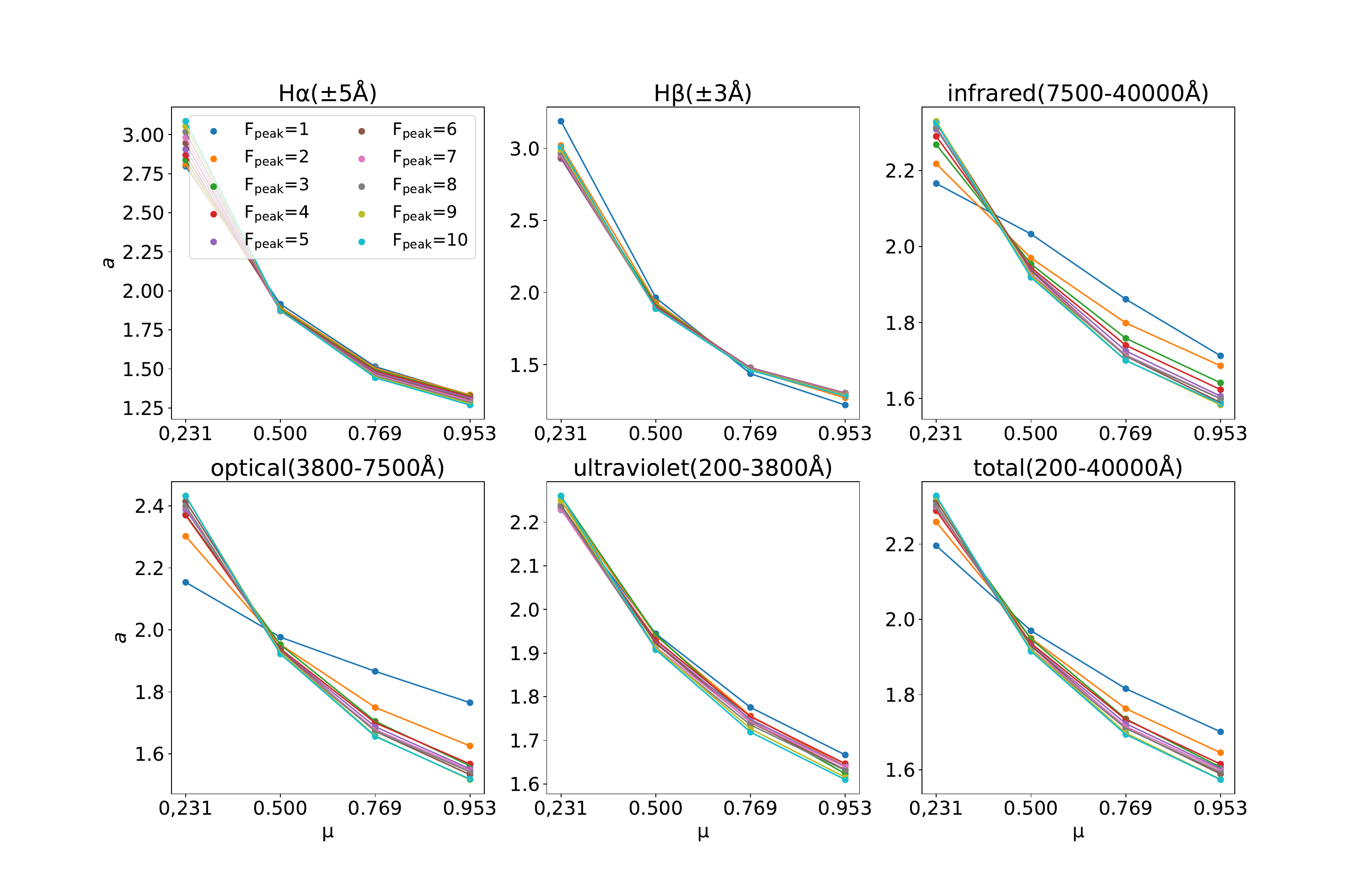}
    \caption{Coefficient $a$ in flux-energy conversion as a function of the flare location with different pass bands. Different colors indicate different flare magnitudes.}
    \label{conversion}
    \end{figure*}

    The next step is to determine the flare radiation energy from $\mathcal{R}$. As is known, a flare radiates energy in all directions. Since there is no "Dyson sphere", it is impossible to detect the flux from a 4$\pi$ solid angle. Therefore, in order to calculate the radiative energy, one should rely on some estimation methods. That is to say, one needs a conversion coefficient $a$ between radiative energy $E$ and irradiance $\mathcal{R}$:
    \begin{equation}
    E_\mathrm{obs}=a\pi D^2\int \Delta \mathcal{R}(\mu,\lambda,t)\mathrm{d}\lambda \mathrm{d}t,
    \label{conversion_obs}
    \end{equation}
    where $\Delta \mathcal{R}(\mu,\lambda,t) = \mathcal{R}(\mu,\lambda,t)-\mathcal{R}(\mu,\lambda,0)$. If the radiation is assumed to be isotropic, the coefficient $a$ is then taken as four. This simple assumption has been employed in many previous works \citep{2024Kowalski,2024Notsu,2011Fletcher,2024Zhao}. However, as discussed in Sect. \ref{Result}, the flare spectra show significant differences if the flare occurs at different locations (e.g., the radiation enhancement at the disk center is greater than that at the disk limb). This indicates that flare radiation is not isotropic, making it inaccurate to choose four as the value of the coefficient. \cite{2006Woods} first considered such anisotropy. By integrating the limb darkening equation over the solid angle which is constrained by observations, they derived a coefficient of 1.4 for optically thick lines. For optically thin lines they assumed isotropy in the hemisphere, giving a coefficient of two. \cite{2010Kretzschmar} also adopted $a=1.4$ in their calculations. However, in our simulation, we can directly calculate the energy due to flares without assumptions as the increase of the outward flux at the flare region surface:
    \begin{equation}
        \Delta L(\lambda,t)=2\pi S_\mathrm{F}\int^1_{0}(I_F(\mu,\lambda,t)-I_F(\mu,\lambda,0))\mu\mathrm{d}\mu.
    \end{equation}
     By integrating $\Delta L(\lambda,t)$ over wavelength and time, we may obtain the total radiated flare energy $E_\mathrm{sim}$:
    \begin{equation}
        E_\mathrm{sim}=\int \Delta L(\lambda,t)\mathrm{d}t\mathrm{d}\lambda.
    \end{equation}
     With both $E_\mathrm{sim}$ (independent of the location) and $\mathcal{R}$ (dependent of the location), we then calculate the coefficient $a$ from Eq. \ref{conversion_obs} directly: 
     \begin{equation}
        a(\mu)=\frac{E_\mathrm{sim}}{\pi D^2\int\Delta\mathcal{R}(\mu,\lambda,t)\mathrm{d}t\mathrm{d}\lambda}.
     \end{equation}
     One should notice that the $a$ is independent of $D$, $R_\odot,$ and $S_\mathrm{F}$. We show the results of $a$ as a function of the location in Fig. \ref{conversion}.
     
     We defined five pass bands: infrared (7500 to 40000 \AA), optical (3800 to 7500 \AA), ultraviolet (200 to 3800 \AA), H$\alpha$ (6562.88 $\pm5 \AA$), H$\beta$ (4861.38 $\pm 3 \AA$), and total (200 to 40000 \AA). 
     The majority of the solar irradiance lies in the wavelength range of 200 to 40000 \AA. The correlation of the coefficient $a$ with $\mu$ for different $F_\mathrm{peak}$ values is quite similar, although the results in weak flares have a large deviation. In each panel, $a$ is negatively correlated with $\mu$, which means that a near-limb flare requires a larger conversion coefficient from flux to energy than a disk-center flare. Except for two Balmer lines, the coefficient varies from about 1.6 to 2.3, with $\mu$ varying from 0.953 to 0.231. For the H$\alpha$ and H$\beta$ lines, the range of variation is greater; the coefficient at the center is 1.25 and can reach 3.0 at the limb, probably because the change in the chromosphere is more remarkable than in the photosphere. When the flare occurs at the disk center, the coefficient for the total pass band lies in the range of (1.58, 1.70), which is a little larger than the value of 1.4 in \cite{2006Woods}. When $\mu$ equals 0.5, the coefficient is two, which is similar to that under the isotropic assumption in the hemisphere. The coefficient value range of different pass bands is listed in Appendix \ref{coefficent table} for reference.

    Such a variation is caused by two effects: the center-to-limb variation of the specific intensity and the projection effect. The former means that the enhancement of the specific intensity $\Delta I_F$ varies with $\mu$, while the latter requires $\Delta I_F$ to be multiplied by $\mu$ (refer to Eq. \ref{deltas_equation}). If we only consider the projection effect, the coefficient is directly the reciprocal of $\mu$, which becomes extremely large when $\mu$ is small. However, the coefficient does not significantly change with $\mu$ due to the compensation of the CLV of specific intensity: $\Delta I_F$ at the limb is larger than at the center, meaning that limb flares are relatively brighter than center flares (see Fig. 4 and Sect. 3.3 in \cite{2010Cheng}). This effect decreases the huge conversion coefficient caused by the small projection area of the limb flare, rendering a gentle change of $a$ (compared to the $1/\mu$) in Fig. \ref{conversion}.

    The flux-energy conversion relationship of a flare is significantly influenced by the flare location. For solar flares, since we can determine the location through imaging observations, it is easy to estimate the radiative energy in each pass band using the coefficient of Fig. \ref{conversion}. We can also determine the energy of a stellar flare if the flare location can be determined (refer to Sect. \ref{Determination of the flare location from observations}). However, the accuracy of $a(\mu)$ still needs to be confirmed by further observations with multiple pass bands.

    \subsection{Energy proportion of different pass bands}

    Stellar flare observations often give a light curve of a certain pass band. To estimate the total radiation energy of a stellar flare, the energy proportion of the pass band in the total wavelength is required. Here, we present the radiation energy proportion across different pass bands in Table \ref{partition}. We note that the values are energy rather than flux, so they are independent of the flare location. The ratio of the optical, infrared, and ultraviolet radiation varies slowly with different $F_\mathrm{peak}$, but it is close to 3:2:5 for most cases. The contribution from the Balmer lines differs as the beam flux changes. For the $F_\mathrm{peak}$=1 case, the H$\alpha$ line contributes to less than 8\% of the optical energy, but for the $F_\mathrm{peak}$=10 case, the contribution drops to 2.6\%. The ratio values in high beam-flux cases are similar to those in previous observations (1.8\% in \cite{2022Namekatac} and 1.5\% in \cite{2022Namekatab}), although they have used the assumption of black-body radiation for the optical continuum, which might underestimate the ratio to some extent \citep{2024Simoes}. We notice that nearly one third of the flare energy comes from the optical pass band, with the majority attributed to the optical continuum. However, caution should be taken in categorizing these flares as observed ``white-light flares'' due to the instrumental sensitivity in real observations. 

    We also point out that the energy proportion is dependent on $E_c$ and $\delta$ since different pass bands of photons arise from different depths of the atmosphere, and the beam penetration depth is strongly influenced by these parameters \citep{2015Allred}. It is also important that the beam parameters are believed to be highly dependent on location within the flare ribbon \citep{2017Druett,2022Osborne,2023Polito}, as mentioned in Sect. \ref{Sun-as-a-star analysis}. The relationship between energy proportion and flare properties remains to be confirmed.
    \begin{table*}
    \caption{Energy partition of each pass band.}
        \centering
        \begin{tabular}{c|c|c|c|c|c|c}
            \hline
            \hline
            $F_\mathrm{peak}$ & H$\alpha$&  H$\beta$&  infrared&  optical& ultraviolet & total energy ($\times 10^{29} \mathrm{erg}$)\\[1ex]
            \hline
            1 &    0.0192&  0.0173&  0.1932&  0.2690& 0.5378 & 0.63\\
            2 &    0.0161&  0.0138&  0.2038&  0.3021& 0.4942 & 1.32\\
            3 &    0.0140&  0.0121&  0.2116&  0.3198& 0.4687 & 2.01\\
            4 &    0.0128&  0.0112&  0.2161&  0.3254& 0.4586 & 2.64\\
            5 &    0.0118&  0.0104&  0.2167&  0.3288& 0.4545 & 3.30\\
            6 &    0.0109&  0.0098&  0.2171&  0.3325& 0.4504 & 3.96\\
            7 &    0.0103&  0.0095&  0.2155&  0.3313& 0.4532 & 4.55\\ 
            8 &    0.0097&  0.0091&  0.2135&  0.3307& 0.4559 & 5.18\\
            9 &    0.0092&  0.0088&  0.2127&  0.3331& 0.4542 & 5.84\\
            10&    0.0087&  0.0084&  0.2081&  0.3315& 0.4605 & 6.54\\
            \hline
        \end{tabular}
        \tablefoot{Total flux of 200 to 40000 \AA\ is unity. The last column shows the total radiation energy.}
        \label{partition}
    \end{table*}

\section{Conclusion}
\label{conclusion}
    In this paper, we present a Sun-as-a-star analysis of simulated solar flares. We begin by calculating the atmospheric response to nonthermal electron beam heating and the corresponding chromospheric lines in a single flare loop. Based on this single loop, we synthesize the Sun-as-a-star profiles using the multi-thread model. Two major parameters -- electron beam flux $F_\mathrm{peak}$ and flare location $\mu$ -- are studied by analyzing the Sun-as-a-star spectra $\Delta S$ of the chromospheric lines. We summarize our results as follows.
    
    \begin{enumerate}
        
        \item The shape of $\Delta S$(H$\alpha$) is characterized by the line enhancement and the line shift. Weak flares show apparent line shift, with an absorption at the blue wings. The line shift signal disappears when the flare occurs at the limb (see Fig. \ref{angle_loop}) or the flare energy is larger (see Fig. \ref{deltas}).

        \item In weak flares, the H$\alpha$ line shows a blueshift, resulting in a red asymmetry, while in strong flares it shows a redshift, leading to a blue asymmetry. The signs of line shift and line asymmetry are then always opposite. Compared to the line shift or the line asymmetry, the time evolution of the equivalent width of H$\alpha$ is better correlated to that of the energy rate, making it a real-time indicator of the flare.

        \item The profile of $\Delta S$ is highly dependent on the flare location. Getting closer to the disk limb, $\Delta S$ becomes wider and shows a more apparent dip in the line core, with multiple peaks at the wings. Two quantities -- $W$ and d$\Delta S$ -- are defined to measure the degree of the line width and the dip, respectively. For both H$\alpha$ and H$\beta$ lines, these two quantities have a stronger dependence on $\mu$ than $F_\mathrm{peak}$. Notably, when $0.2<\mu<0.4$, d$\Delta S$ is primarily dependent on the location. We recommend using these two quantities to diagnose the location and magnitude of the stellar flare.

        \item One needs a conversion coefficient $a$ to calculate the radiation energy from the observed flux (Eq. \ref{conversion_obs}). We show that $a$ is a function of the location $\mu$. Additionally, the radiation energy proportion of different pass bands is presented in Table \ref{partition} for reference.

    \end{enumerate}

    The above results in our simulations generally agree with previous observations. Yet, we will rely on more future observations, such as the full-disk observation of the H$\alpha$ line by the Chinese H$\alpha$ Solar Explorer \citep{2019Li},  to verify the models and diagnostics. On the other hand, simulations of other transient events including those by \cite{2022Yang} and \cite{2024Ikuta} are needed to distinguish these events from flares in the observations. 

 \begin{acknowledgements}
 We thank the referee, Malcolm Druett, for his detailed and valuable suggestions that have greatly improved this paper.
       This project has been funded by National Key R\&D Program of China under grant 2022YFF0503004 and by NSFC under grant 12127901. J.H. is also funded by the European Union through the European Research Council (ERC) under the Horizon Europe program (MAGHEAT, grant agreement 101088184). The Institute for Solar Physics is supported by a grant for research infrastructures of national importance from the Swedish Research Council (registration number 2021-00169).
 \end{acknowledgements}

\bibliographystyle{aa}
\bibliography{citation}

\begin{thebibliography}{66}
\expandafter\ifx\csname natexlab\endcsname\relax\def\natexlab#1{#1}\fi

\bibitem[{{Allred} {et~al.}(2015){Allred}, {Kowalski}, \&
  {Carlsson}}]{2015Allred}
{Allred}, J.~C., {Kowalski}, A.~F., \& {Carlsson}, M. 2015, \apj, 809, 104

\bibitem[{{Amazo-G{\'o}mez} {et~al.}(2023){Amazo-G{\'o}mez},
  {Alvarado-G{\'o}mez}, {Poppenh{\"a}ger}, {Hussain}, {Wood}, {Drake}, {do
  Nascimento}, {Anthony}, {Sanz-Forcada}, {Stelzer}, {Del Sordo}, {Damasso},
  {Redfield}, {Donati}, {K{\"o}nig}, {H{\'e}brard}, \&
  {Miles-P{\'a}ez}}]{2023Amazo-Gomez}
{Amazo-G{\'o}mez}, E.~M., {Alvarado-G{\'o}mez}, J.~D., {Poppenh{\"a}ger}, K.,
  {et~al.} 2023, \mnras, 524, 5725

\bibitem[{{Aschwanden} \& {Alexander}(2001)}]{2001Aschwanden}
{Aschwanden}, M.~J. \& {Alexander}, D. 2001, \solphys, 204, 91

\bibitem[{{Boisse} {et~al.}(2009){Boisse}, {Moutou}, {Vidal-Madjar}, {Bouchy},
  {Pont}, {H{\'e}brard}, {Bonfils}, {Croll}, {Delfosse}, {Desort}, {Forveille},
  {Lagrange}, {Loeillet}, {Lovis}, {Matthews}, {Mayor}, {Pepe}, {Perrier},
  {Queloz}, {Rowe}, {Santos}, {S{\'e}gransan}, \& {Udry}}]{2009Boisse}
{Boisse}, I., {Moutou}, C., {Vidal-Madjar}, A., {et~al.} 2009, \aap, 495, 959

\bibitem[{{Canfield} {et~al.}(1990){Canfield}, {Penn}, {Wulser}, \&
  {Kiplinger}}]{1990Canfield}
{Canfield}, R.~C., {Penn}, M.~J., {Wulser}, J.-P., \& {Kiplinger}, A.~L. 1990,
  \apj, 363, 318

\bibitem[{{Carlsson} \& {Stein}(1992)}]{1992ApJ...397L..59C}
{Carlsson}, M. \& {Stein}, R.~F. 1992, \apjl, 397, L59

\bibitem[{{Carlsson} \& {Stein}(1995)}]{1995ApJ...440L..29C}
{Carlsson}, M. \& {Stein}, R.~F. 1995, \apjl, 440, L29

\bibitem[{{Carlsson} \& {Stein}(1997)}]{1997ApJ...481..500C}
{Carlsson}, M. \& {Stein}, R.~F. 1997, \apj, 481, 500

\bibitem[{{Carlsson} \& {Stein}(2002)}]{2002ApJ...572..626C}
{Carlsson}, M. \& {Stein}, R.~F. 2002, \apj, 572, 626

\bibitem[{{Cheng} {et~al.}(2010){Cheng}, {Ding}, \& {Carlsson}}]{2010Cheng}
{Cheng}, J.~X., {Ding}, M.~D., \& {Carlsson}, M. 2010, \apj, 711, 185

\bibitem[{{Criscuoli} {et~al.}(2023){Criscuoli}, {Marchenko}, {DeLand},
  {Choudhary}, \& {Kopp}}]{2023Criscuoli}
{Criscuoli}, S., {Marchenko}, S., {DeLand}, M., {Choudhary}, D., \& {Kopp}, G.
  2023, \apj, 951, 151

\bibitem[{{Druett} {et~al.}(2024){Druett}, {Ruan}, \& {Keppens}}]{2024druett}
{Druett}, M., {Ruan}, W., \& {Keppens}, R. 2024, \aap, 684, A171

\bibitem[{{Druett} {et~al.}(2017){Druett}, {Scullion}, {Zharkova}, {Matthews},
  {Zharkov}, \& {Rouppe van der Voort}}]{2017Druett}
{Druett}, M., {Scullion}, E., {Zharkova}, V., {et~al.} 2017, Nature
  Communications, 8, 15905

\bibitem[{{Druett} \& {Zharkova}(2018)}]{2018Druett}
{Druett}, M.~K. \& {Zharkova}, V.~V. 2018, \aap, 610, A68

\bibitem[{{Fletcher} {et~al.}(2011){Fletcher}, {Dennis}, {Hudson}, {Krucker},
  {Phillips}, {Veronig}, {Battaglia}, {Bone}, {Caspi}, {Chen}, {Gallagher},
  {Grigis}, {Ji}, {Liu}, {Milligan}, \& {Temmer}}]{2011Fletcher}
{Fletcher}, L., {Dennis}, B.~R., {Hudson}, H.~S., {et~al.} 2011, \ssr, 159, 19

\bibitem[{{Haisch} {et~al.}(1991){Haisch}, {Strong}, \& {Rodono}}]{1991Haisch}
{Haisch}, B., {Strong}, K.~T., \& {Rodono}, M. 1991, \araa, 29, 275

\bibitem[{{Hong} {et~al.}(2022){Hong}, {Carlsson}, \& {Ding}}]{2022Hong}
{Hong}, J., {Carlsson}, M., \& {Ding}, M.~D. 2022, \aap, 661, A77

\bibitem[{{Hong} {et~al.}(2019){Hong}, {Li}, {Ding}, \& {Carlsson}}]{2019Hong}
{Hong}, J., {Li}, Y., {Ding}, M.~D., \& {Carlsson}, M. 2019, \apj, 879, 128

\bibitem[{{Ikuta} \& {Shibata}(2024)}]{2024Ikuta}
{Ikuta}, K. \& {Shibata}, K. 2024, \apj, 963, 50

\bibitem[{{Kerr} {et~al.}(2019){Kerr}, {Carlsson}, {Allred}, {Young}, \&
  {Daw}}]{2019ApJ...871...23K}
{Kerr}, G.~S., {Carlsson}, M., {Allred}, J.~C., {Young}, P.~R., \& {Daw}, A.~N.
  2019, \apj, 871, 23

\bibitem[{{Kowalski}(2024)}]{2024Kowalski}
{Kowalski}, A.~F. 2024, Living Reviews in Solar Physics, 21, 1

\bibitem[{{Kowalski} {et~al.}(2017){Kowalski}, {Allred}, {Uitenbroek},
  {Tremblay}, {Brown}, {Carlsson}, {Osten}, {Wisniewski}, \&
  {Hawley}}]{2017Kowalski}
{Kowalski}, A.~F., {Allred}, J.~C., {Uitenbroek}, H., {et~al.} 2017, \apj, 837,
  125

\bibitem[{{Kretzschmar}(2011)}]{2011Ketzschmar}
{Kretzschmar}, M. 2011, \aap, 530, A84

\bibitem[{{Kretzschmar} {et~al.}(2010){Kretzschmar}, {de Wit}, {Schmutz},
  {Mekaoui}, {Hochedez}, \& {Dewitte}}]{2010Kretzschmar}
{Kretzschmar}, M., {de Wit}, T.~D., {Schmutz}, W., {et~al.} 2010, Nature
  Physics, 6, 690

\bibitem[{{Kuridze} {et~al.}(2015){Kuridze}, {Mathioudakis}, {Sim{\~o}es},
  {Rouppe van der Voort}, {Carlsson}, {Jafarzadeh}, {Allred}, {Kowalski},
  {Kennedy}, {Fletcher}, {Graham}, \& {Keenan}}]{2015Kuridze}
{Kuridze}, D., {Mathioudakis}, M., {Sim{\~o}es}, P.~J.~A., {et~al.} 2015, \apj,
  813, 125

\bibitem[{{Li} {et~al.}(2019){Li}, {Ding}, {Hong}, {Li}, \& {Gan}}]{2019Li}
{Li}, Y., {Ding}, M.~D., {Hong}, J., {Li}, H., \& {Gan}, W.~Q. 2019, \apj, 879,
  30

\bibitem[{{Livingston} {et~al.}(2007){Livingston}, {Wallace}, {White}, \&
  {Giampapa}}]{2007Livingston}
{Livingston}, W., {Wallace}, L., {White}, O.~R., \& {Giampapa}, M.~S. 2007,
  \apj, 657, 1137

\bibitem[{{Maehara} {et~al.}(2012){Maehara}, {Shibayama}, {Notsu}, {Notsu},
  {Nagao}, {Kusaba}, {Honda}, {Nogami}, \& {Shibata}}]{2012Maehara}
{Maehara}, H., {Shibayama}, T., {Notsu}, S., {et~al.} 2012, \nat, 485, 478

\bibitem[{{Melbourne} {et~al.}(2020){Melbourne}, {Youngblood}, {France},
  {Froning}, {Pineda}, {Shkolnik}, {Wilson}, {Wood}, {Basu}, {Roberge},
  {Schlieder}, {Cauley}, {Loyd}, {Newton}, {Schneider}, {Arulanantham},
  {Berta-Thompson}, {Brown}, {Buccino}, {Kempton}, {Linsky}, {Logsdon},
  {Mauas}, {Pagano}, {Peacock}, {Redfield}, {Rugheimer}, {Schneider}, {Teal},
  {Tian}, {Tilipman}, \& {Vieytes}}]{2020Melbourne}
{Melbourne}, K., {Youngblood}, A., {France}, K., {et~al.} 2020, \aj, 160, 269

\bibitem[{{Namekata} {et~al.}(2022{\natexlab{a}}){Namekata}, {Ichimoto},
  {Ishii}, \& {Shibata}}]{2022Namekatac}
{Namekata}, K., {Ichimoto}, K., {Ishii}, T.~T., \& {Shibata}, K.
  2022{\natexlab{a}}, \apj, 933, 209

\bibitem[{{Namekata} {et~al.}(2022{\natexlab{b}}){Namekata}, {Maehara},
  {Honda}, {Notsu}, {Okamoto}, {Takahashi}, {Takayama}, {Ohshima}, {Saito},
  {Katoh}, {Tozuka}, {Murata}, {Ogawa}, {Niwano}, {Adachi}, {Oeda},
  {Shiraishi}, {Isogai}, {Nogami}, \& {Shibata}}]{2022Namekatab}
{Namekata}, K., {Maehara}, H., {Honda}, S., {et~al.} 2022{\natexlab{b}}, \apjl,
  926, L5

\bibitem[{{Namekata} {et~al.}(2021){Namekata}, {Maehara}, {Honda}, {Notsu},
  {Okamoto}, {Takahashi}, {Takayama}, {Ohshima}, {Saito}, {Katoh}, {Tozuka},
  {Murata}, {Ogawa}, {Niwano}, {Adachi}, {Oeda}, {Shiraishi}, {Isogai}, {Seki},
  {Ishii}, {Ichimoto}, {Nogami}, \& {Shibata}}]{2021Namekata}
{Namekata}, K., {Maehara}, H., {Honda}, S., {et~al.} 2021, Nature Astronomy, 6,
  241

\bibitem[{{Neupert}(1968)}]{1968Neupert}
{Neupert}, W.~M. 1968, \apjl, 153, L59

\bibitem[{{Notsu} {et~al.}(2024){Notsu}, {Kowalski}, {Maehara}, {Namekata},
  {Hamaguchi}, {Enoto}, {Tristan}, {Hawley}, {Davenport}, {Honda}, {Ikuta},
  {Inoue}, {Namizaki}, {Nogami}, \& {Shibata}}]{2024Notsu}
{Notsu}, Y., {Kowalski}, A.~F., {Maehara}, H., {et~al.} 2024, \apj, 961, 189

\bibitem[{{Notsu} {et~al.}(2013){Notsu}, {Shibayama}, {Maehara}, {Notsu},
  {Nagao}, {Honda}, {Ishii}, {Nogami}, \& {Shibata}}]{2013Notsu}
{Notsu}, Y., {Shibayama}, T., {Maehara}, H., {et~al.} 2013, \apj, 771, 127

\bibitem[{{Osborne} \& {Fletcher}(2022)}]{2022Osborne}
{Osborne}, C. M.~J. \& {Fletcher}, L. 2022, \mnras, 516, 6066

\bibitem[{{Otsu} \& {Asai}(2024)}]{2024Otsu}
{Otsu}, T. \& {Asai}, A. 2024, \apj, 964, 75

\bibitem[{{Otsu} {et~al.}(2022){Otsu}, {Asai}, {Ichimoto}, {Ishii}, \&
  {Namekata}}]{2022Otsu}
{Otsu}, T., {Asai}, A., {Ichimoto}, K., {Ishii}, T.~T., \& {Namekata}, K. 2022,
  \apj, 939, 98

\bibitem[{{Parker}(1963)}]{1963Parker}
{Parker}, E.~N. 1963, \apjs, 8, 177

\bibitem[{{Pereira} \& {Uitenbroek}(2015)}]{2015Pereira}
{Pereira}, T. M.~D. \& {Uitenbroek}, H. 2015, \aap, 574, A3

\bibitem[{{Pietrow} {et~al.}(2024{\natexlab{a}}){Pietrow}, {Cretignier},
  {Druett}, {Alvarado-G{\'o}mez}, {Hofmeister}, {Verma}, {Kamlah}, {Baratella},
  {Amazo-G{\'o}mez}, {Kontogiannis}, {Dineva}, {Warmuth}, {Denker},
  {Poppenhaeger}, {Andriienko}, {Dumusque}, \& {L{\"o}fdahl}}]{2024Pietrow}
{Pietrow}, A.~G.~M., {Cretignier}, M., {Druett}, M.~K., {et~al.}
  2024{\natexlab{a}}, \aap, 682, A46

\bibitem[{{Pietrow} {et~al.}(2024{\natexlab{b}}){Pietrow}, {Druett}, \&
  {Singh}}]{2024Pietrowb}
{Pietrow}, A.~G.~M., {Druett}, M.~K., \& {Singh}, V. 2024{\natexlab{b}}, \aap,
  685, A137

\bibitem[{{Pietrow} {et~al.}(2023){Pietrow}, {Hoppe}, {Bergemann}, \&
  {Calvo}}]{2023Pietrowa}
{Pietrow}, A.~G.~M., {Hoppe}, R., {Bergemann}, M., \& {Calvo}, F. 2023, \aap,
  672, L6

\bibitem[{{Pietrow} \& {Pastor Yabar}(2024)}]{2024Pietrowc}
{Pietrow}, A. G.~M. \& {Pastor Yabar}, A. 2024, in IAU Symposium, Vol. 365, IAU
  Symposium, ed. A.~V. {Getling} \& L.~L. {Kitchatinov}, 389--393

\bibitem[{{Pinto} {et~al.}(2015){Pinto}, {Vilmer}, \& {Brun}}]{2015Pinto}
{Pinto}, R.~F., {Vilmer}, N., \& {Brun}, A.~S. 2015, \aap, 576, A37

\bibitem[{{Polito} {et~al.}(2023){Polito}, {Kerr}, {Xu}, {Sadykov}, \&
  {Lorincik}}]{2023Polito}
{Polito}, V., {Kerr}, G.~S., {Xu}, Y., {Sadykov}, V.~M., \& {Lorincik}, J.
  2023, \apj, 944, 104

\bibitem[{{Pulkkinen}(2007)}]{2007Pulkkinen}
{Pulkkinen}, T. 2007, Living Reviews in Solar Physics, 4, 1

\bibitem[{{Qiu} \& {Longcope}(2016)}]{2016Qiu}
{Qiu}, J. \& {Longcope}, D.~W. 2016, \apj, 820, 14

\bibitem[{{Ruan} {et~al.}(2024){Ruan}, {Keppens}, {Yan}, \&
  {Antolin}}]{2024Ruan}
{Ruan}, W., {Keppens}, R., {Yan}, L., \& {Antolin}, P. 2024, \apj, 967, 82

\bibitem[{{Rubio da Costa} {et~al.}(2016){Rubio da Costa}, {Kleint},
  {Petrosian}, {Liu}, \& {Allred}}]{2016Costa}
{Rubio da Costa}, F., {Kleint}, L., {Petrosian}, V., {Liu}, W., \& {Allred},
  J.~C. 2016, \apj, 827, 38

\bibitem[{{Shibata} \& {Magara}(2011)}]{2011Shibata}
{Shibata}, K. \& {Magara}, T. 2011, Living Reviews in Solar Physics, 8, 6

\bibitem[{{Shibata} \& {Tanuma}(2001)}]{2001Shibata}
{Shibata}, K. \& {Tanuma}, S. 2001, Earth, Planets and Space, 53, 473

\bibitem[{{Sim{\~o}es} {et~al.}(2024){Sim{\~o}es}, {Ara{\'u}jo}, {V{\'a}lio},
  \& {Fletcher}}]{2024Simoes}
{Sim{\~o}es}, P. J.~A., {Ara{\'u}jo}, A., {V{\'a}lio}, A., \& {Fletcher}, L.
  2024, \mnras, 528, 2562

\bibitem[{{Sim{\~o}es} {et~al.}(2017){Sim{\~o}es}, {Kerr}, {Fletcher},
  {Hudson}, {Gim{\'e}nez de Castro}, \& {Penn}}]{2017Simoes}
{Sim{\~o}es}, P. J.~A., {Kerr}, G.~S., {Fletcher}, L., {et~al.} 2017, \aap,
  605, A125

\bibitem[{{Tsurutani} {et~al.}(2003){Tsurutani}, {Gonzalez}, {Lakhina}, \&
  {Alex}}]{2003Tsurtani}
{Tsurutani}, B.~T., {Gonzalez}, W.~D., {Lakhina}, G.~S., \& {Alex}, S. 2003,
  Journal of Geophysical Research (Space Physics), 108, 1268

\bibitem[{{Vaughan} {et~al.}(1978){Vaughan}, {Preston}, \&
  {Wilson}}]{1978Vaughan}
{Vaughan}, A.~H., {Preston}, G.~W., \& {Wilson}, O.~C. 1978, \pasp, 90, 267

\bibitem[{{Warren}(2006)}]{2006Warren}
{Warren}, H.~P. 2006, \apj, 637, 522

\bibitem[{{Woods} {et~al.}(2004){Woods}, {Eparvier}, {Fontenla}, {Harder},
  {Kopp}, {McClintock}, {Rottman}, {Smiley}, \& {Snow}}]{2004Woods}
{Woods}, T.~N., {Eparvier}, F.~G., {Fontenla}, J., {et~al.} 2004, \grl, 31,
  L10802

\bibitem[{{Woods} {et~al.}(2006){Woods}, {Kopp}, \& {Chamberlin}}]{2006Woods}
{Woods}, T.~N., {Kopp}, G., \& {Chamberlin}, P.~C. 2006, Journal of Geophysical
  Research (Space Physics), 111, A10S14

\bibitem[{{Wu} {et~al.}(2022){Wu}, {Chen}, {Tian}, {Zhang}, {Shi}, {He}, {Lu},
  {Xu}, \& {Wang}}]{2022Wu}
{Wu}, Y., {Chen}, H., {Tian}, H., {et~al.} 2022, \apj, 928, 180

\bibitem[{{Xu} {et~al.}(2022){Xu}, {Tian}, {Hou}, {Yang}, {Gao}, \&
  {Bai}}]{2022Xu}
{Xu}, Y., {Tian}, H., {Hou}, Z., {et~al.} 2022, \apj, 931, 76

\bibitem[{{Yang} {et~al.}(2022){Yang}, {Tian}, {Bai}, {Chen}, {Guo}, {Zhu},
  {Cheng}, {Gao}, {Xu}, {Chen}, \& {Zhang}}]{2022Yang}
{Yang}, Z., {Tian}, H., {Bai}, X., {et~al.} 2022, \apjs, 260, 36

\bibitem[{{Yu} {et~al.}(2023){Yu}, {Hong}, \& {Ding}}]{2023Yu}
{Yu}, H.~C., {Hong}, J., \& {Ding}, M.~D. 2023, \aap, 675, A171

\bibitem[{{Zhao} {et~al.}(2024){Zhao}, {Hua}, {Cheng}, {Li}, \&
  {Ding}}]{2024Zhao}
{Zhao}, Z.~H., {Hua}, Z.~Q., {Cheng}, X., {Li}, Z.~Y., \& {Ding}, M.~D. 2024,
  \apj, 961, 130

\bibitem[{{Zhou} {et~al.}(2022){Zhou}, {Hong}, {Li}, \& {Ding}}]{2022zhou}
{Zhou}, Y.-A., {Hong}, J., {Li}, Y., \& {Ding}, M.~D. 2022, \apj, 926, 223

\bibitem[{{Zills} {et~al.}(2024){Zills}, {Criscuoli}, {Bertello}, \&
  {Pevtsov}}]{2024Zills}
{Zills}, G., {Criscuoli}, S., {Bertello}, L., \& {Pevtsov}, A. 2024, Frontiers
  in Astronomy and Space Sciences, 10, 1328364

\end{thebibliography}


\begin{appendix}
\onecolumn
    \section{Additional table for Fig. \ref{conversion}}
    \label{coefficent table}
        \begin{table}[h!]
        \centering
        \caption{Values of conversion coefficient $a$.}
        \begin{tabular}{c|cccccc}  
        \hline
        \hline
        
            flare location($\mu$)&  H$\alpha$&  H$\beta$&  infrared&  optical& ultraviolet&  total\\ [1ex]
            \hline
            
            0.231 &  [2.84, 3.13]&  [2.97, 3.24]&  [2.16, 2.35]&  [2.15, 2.45]&  [2.26, 2.29]& [2.21, 2.35]\\
            0.500 &  [1.90, 1.94]&  [1.92, 1.99]&  [1.93, 2.02]&  [1.93, 1.96]&  [1.93, 1.97]& [1.93, 1.98]\\
            0.769 &  [1.46, 1.54]&  [1.46, 1.50]&  [1.71, 1.84]&  [1.66, 1.85]&  [1.74, 1.80]& [1.71, 1.82]\\
            0.953 &  [1.29, 1.35]&  [1.24, 1.32]&  [1.59, 1.69]&  [1.52, 1.74]&  [1.63, 1.69]& [1.58, 1.70]\\
            \hline
        \end{tabular}
        \tablefoot{Supplement to Fig. \ref{conversion}. The range of values is taken from the minimum and maximum coefficients of 10 flare cases.}
        \end{table}

    \section{Additional figures}
    \label{additional figures}
        \begin{figure}[h!]
        \centering
        \includegraphics[scale=0.38]{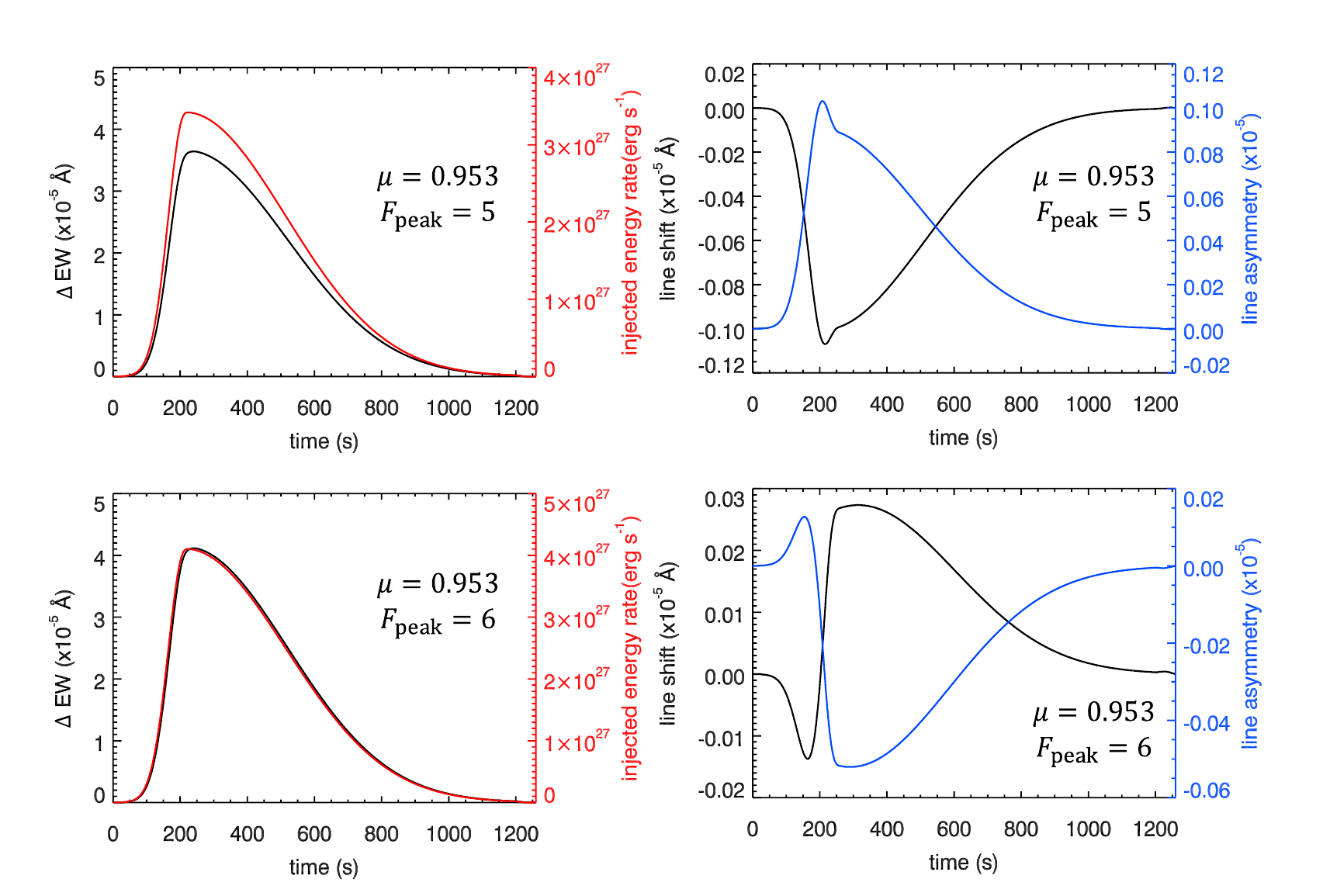}
        \caption{Same as Fig.~\ref{profile} but $F_\mathrm{peak}=5$ and 6.}
        \label{appendix1}
        \end{figure}
    
        \begin{figure}[h!]
        \centering
        \includegraphics[scale=0.38]{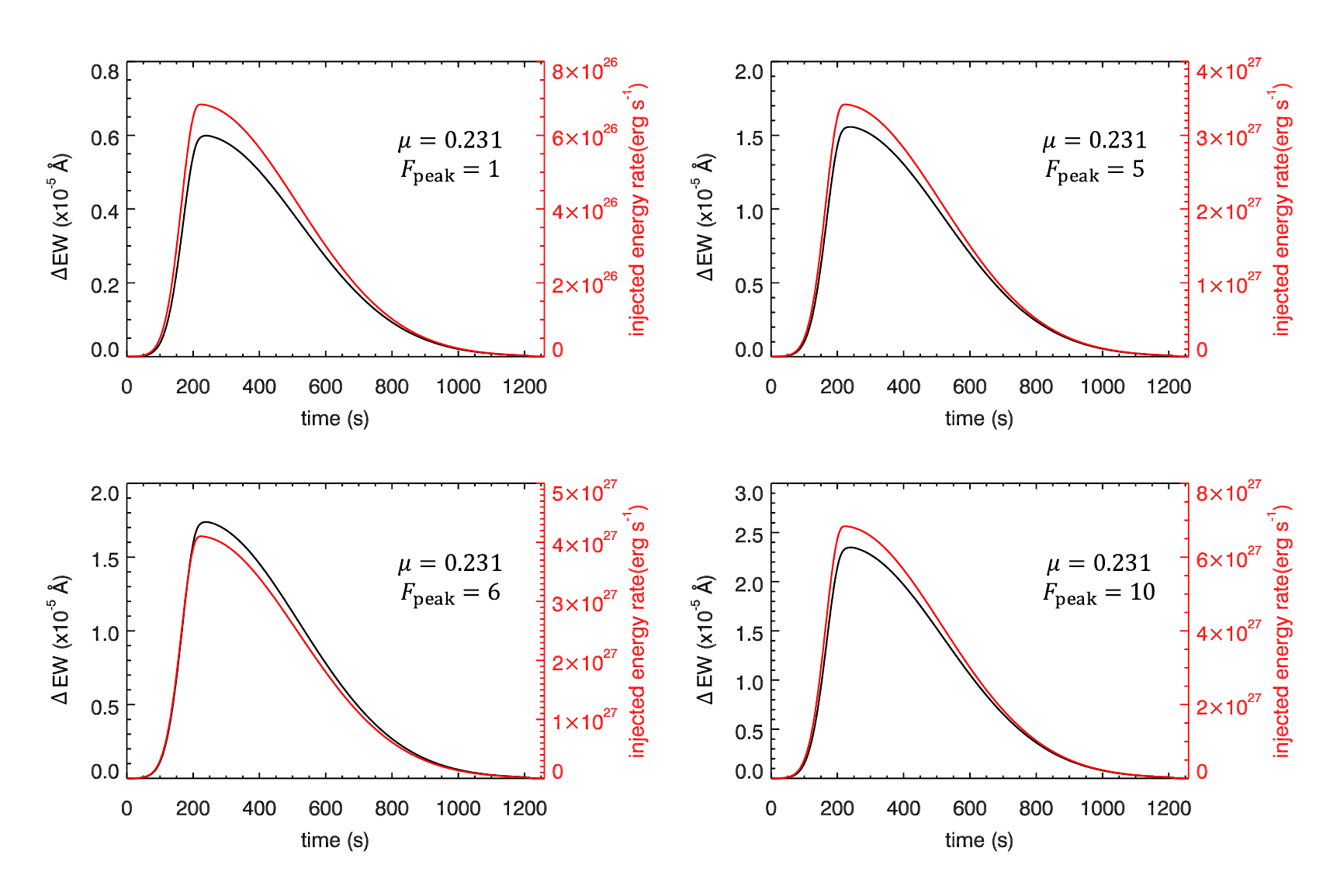}
        \caption{Injected energy rate of the non-thermal electron beam (red line) and equivalent width (black line) when flares occur at disk limb.} 
        \label{appendix2}
        \end{figure}
\end{appendix}

\end{document}